\documentclass[%
 aps,
 prl,
 amsmath,amssymb,
 reprint,%
]{revtex4-2}

\usepackage{graphicx}%
\usepackage[utf8]{inputenc}
\usepackage[T1]{fontenc}
\usepackage{hyperref}
\usepackage{tikz}
\usepackage{microtype}

\begin{document}

\preprint{AIP/123-QED}

\title[]{Connecting minimal chimeras and fully asymmetric chaotic attractors through equivariant pitchfork bifurcations}

\author{Sindre W. Haugland}
\author{Katharina Krischer}
 \email{krischer@tum.de.}
\affiliation{Physics Department, Nonequilibrium Chemical Physics, Technical University of Munich,
  James-Franck-Str. 1, D-85748 Garching, Germany}

\date{\today}%

\begin{abstract}
Highly symmetric networks %
can exhibit partly symmetry-broken 
states, including clusters and %
chimera states, i.e., states of coexisting synchronized and unsynchronized elements.
We address the $\mathbb{S}_4$ %
permutation symmetry %
of four globally coupled Stuart-Landau oscillators and uncover %
an interconnected web of differently symmetric solutions.
Among these are chaotic $2\!-\!1\!-\!1$ minimal chimeras %
that arise from $2\!-\!1\!-\!1$ periodic solutions in a period-doubling cascade, as well as fully asymmetric chaotic states arising similarly from periodic $1\!-\!1\!-\!1\!-\!1$ solutions.
A backbone %
of equivariant pitchfork bifurcations mediate between the two %
cascades, culminating in equivariant pitchforks of chaotic attractors.
\end{abstract}

\maketitle

Symmetrically coupled identical units do not 
always behave identically.
Instead, they may conglomerate into several distinct internally synchronized clusters~\cite{Kaneko_PhysicaD_1990,Golomb_PRA_1992,Hakim_PRA_1992,Okuda_PhysicaD_1993}.
Or only some of them may synchronize, while the others remain unclustered~\cite{Kaneko_PhysicaD_1990,Nakagawa_PTPS_1993,Kuramoto_NPCS_2002}, forming a so-called chimera state~\cite{Abrams_PRL_2004}.
The latter phenomenon in particular has garnered a lot of attention over the last two decades and has been identified in a wide range of theoretical models and experimental setups~\cite{Panaggio_Nonlinearity_2015,Scholl_EPJST_2016,OmelChenko_Nonlinearity_2018}.

A relatively recent question related to chimera states is that of small or minimal chimeras, %
identified
in ensembles of only a few~\cite{Ashwin_Chaos_2015}, usually four~\cite{Panaggio_PRE_2016,Hart_Chaos_2016,Dudkowski_SciRep_2016,Rohm_PRE_2016,Kemeth_PRL_2018}, sometimes three~\cite{Wojewoda_SciRep_2016,Maistrenko_PRE_2017} oscillators.
Some of these, like some of the macroscopic chimeras previously identified~\cite{Kaneko_PhysicaD_1990,Nakagawa_PTPS_1993,Schmidt_Chaos_2014,Sethia_PRL_2014,Bohm_PRE_2015}, even occur in the case of global coupling, that is, when all the oscillators are coupled equally to each other~\cite{Rohm_PRE_2016,Kemeth_PRL_2018}.
The corresponding models have in common that they do not consist of pure phase oscillators, but of oscillators with several components.

In this Letter, we %
use a
model of four globally coupled complex-valued oscillators to demonstrate how chaotic minimal chimera states arise as a consequence of the stepwise breaking of the symmetry of the underlying equations.
Our analysis uncovers a %
network
of interconnected bifurcation curves, wherein the chimeras are embedded together with other partially synchronized states.
The backbone of this network is a sequence of equivariant pitchfork bifurcations~\cite{Moehlis_ScholPed_2007}, on the one side of which two of the oscillators are clustered, while on the other, none of them are.
Along this equivariant pitchfork sequence, several other bifurcations affect the symmetries and the periodicities of the stable solutions on either side, culminating in two interconnected period-doubling cascades to chaos. 
In the chaotic regime of this cascade, equivariant pitchforks of chaotic attractors govern the transition between chimeras and asymmetric chaos, allowing for full-circle transitions via one route from regular motion to chaos and back again via a different route. 

\begin{figure*}[ht]
    \centering
    \begin{tikzpicture}
        \node (fig1) at (0,0)
        {\centering
    \begin{minipage}{1.98\columnwidth}
    \centering
    \vspace{-1.5mm}
    \hspace{-4mm}
    \includegraphics[width=0.250\columnwidth]{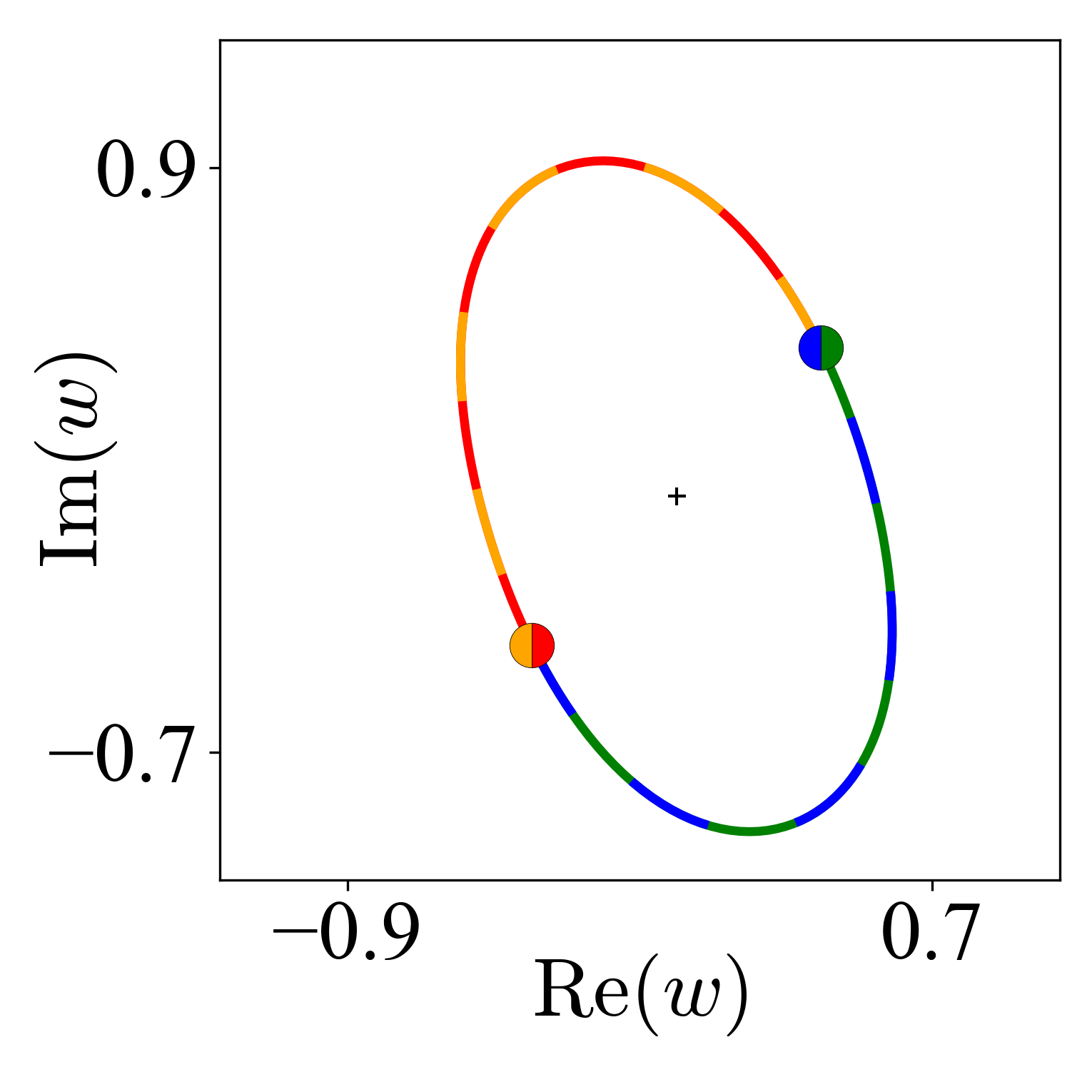}
    \hspace{-2.5mm}          
    \includegraphics[width=0.250\columnwidth]{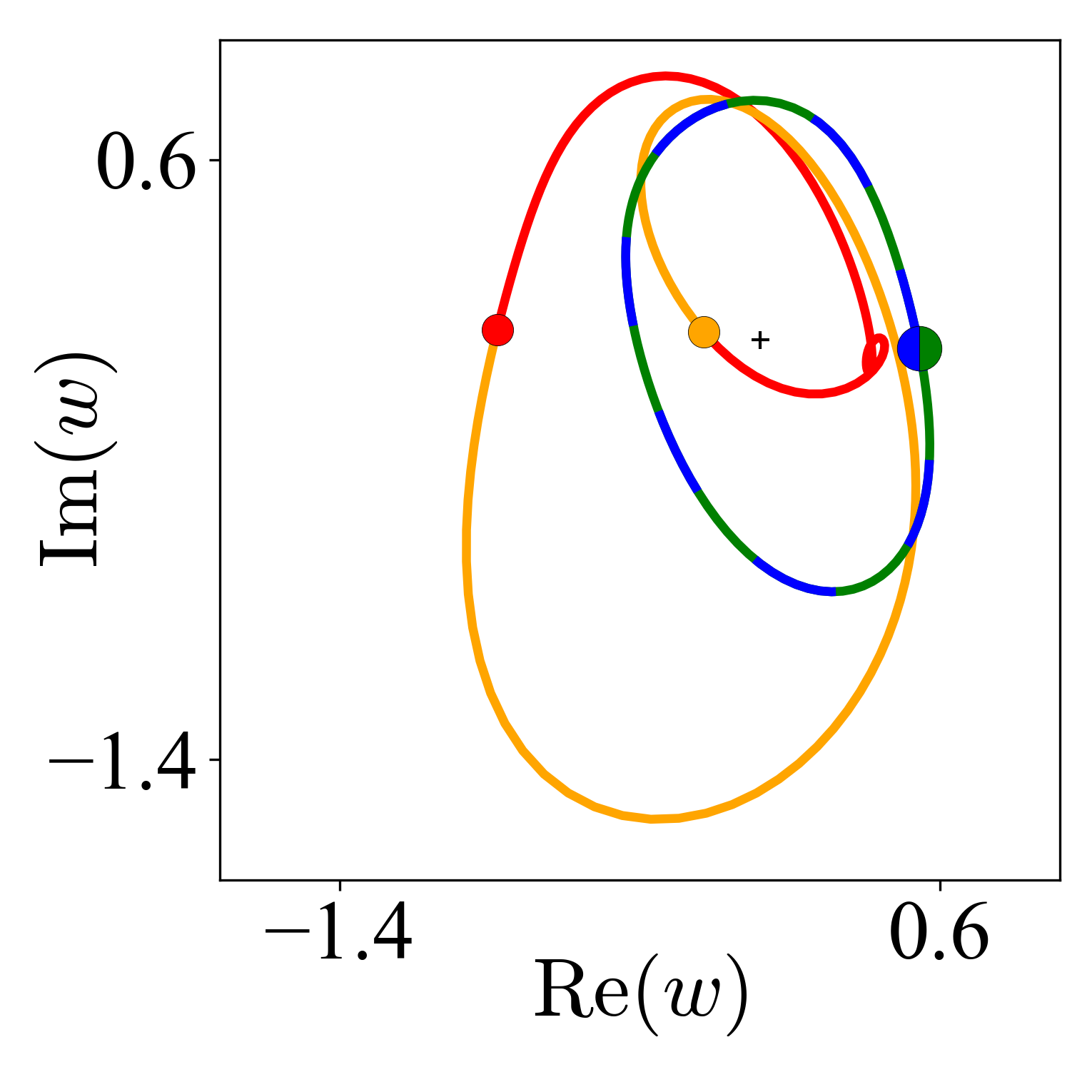}
    \hspace{-2.5mm}          
    \includegraphics[width=0.250\columnwidth]{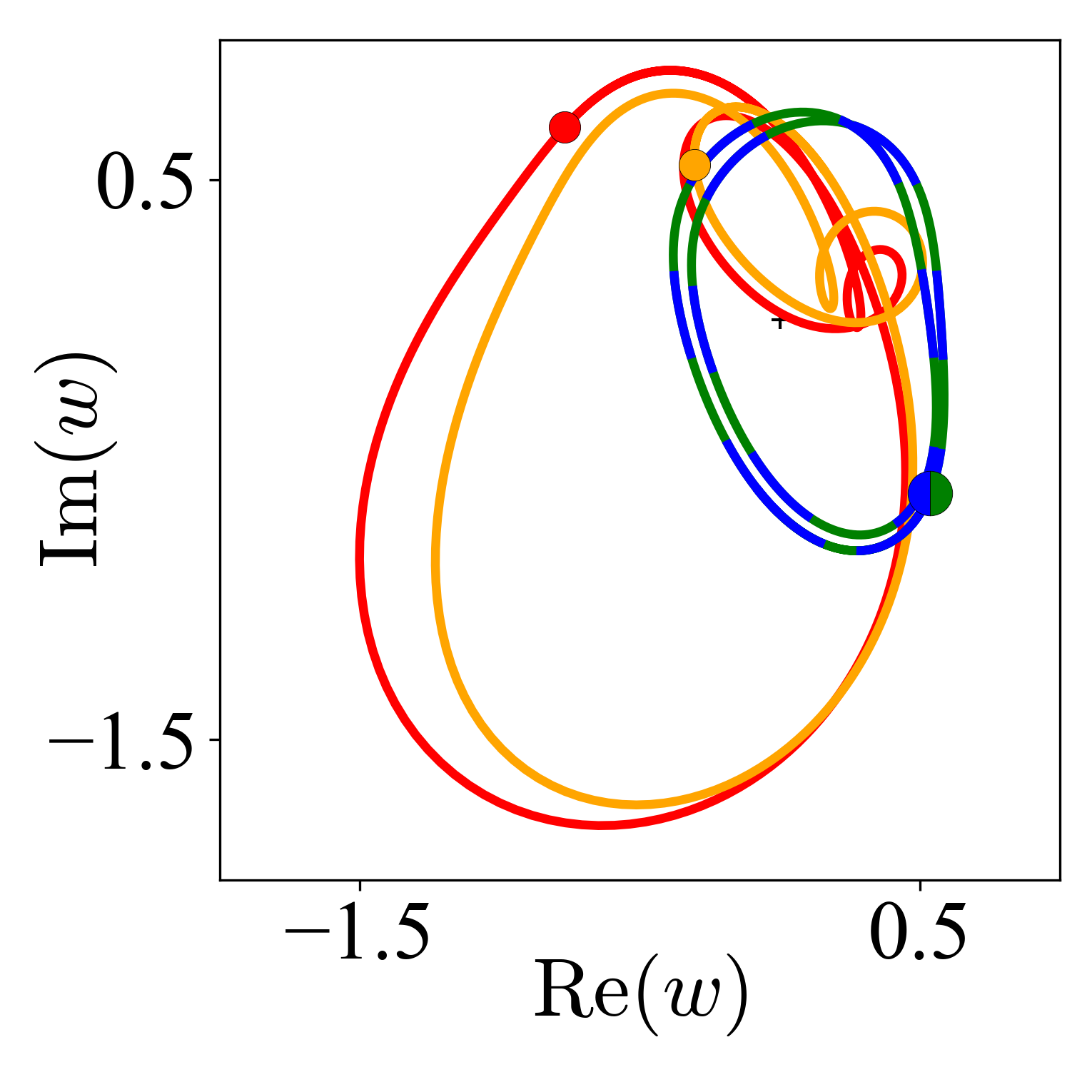}
    \hspace{-2.5mm}          
    \includegraphics[width=0.250\columnwidth]{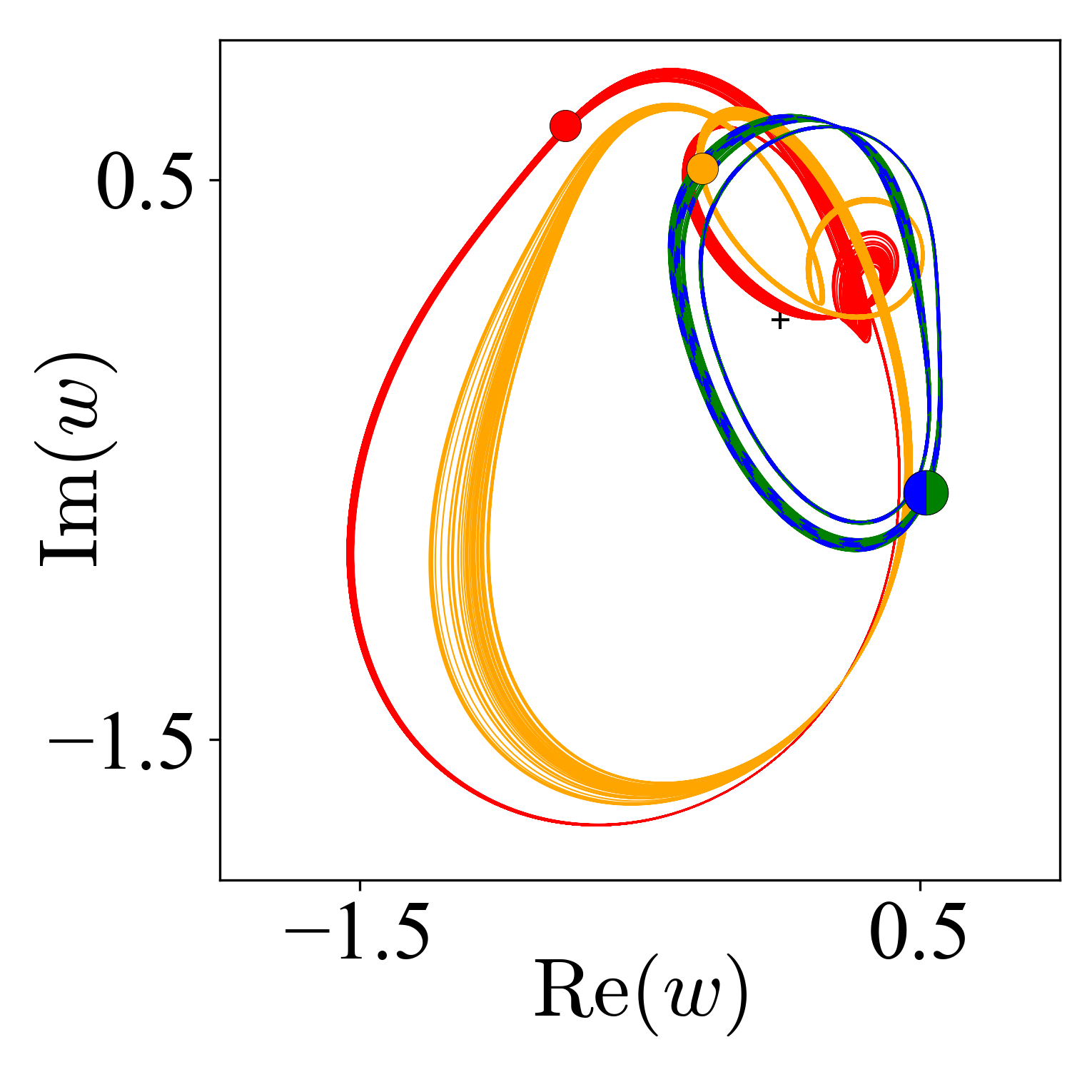}\\
    \vspace{-1.0mm}          
    \hspace{-3mm}            
    \includegraphics[width=0.250\columnwidth]{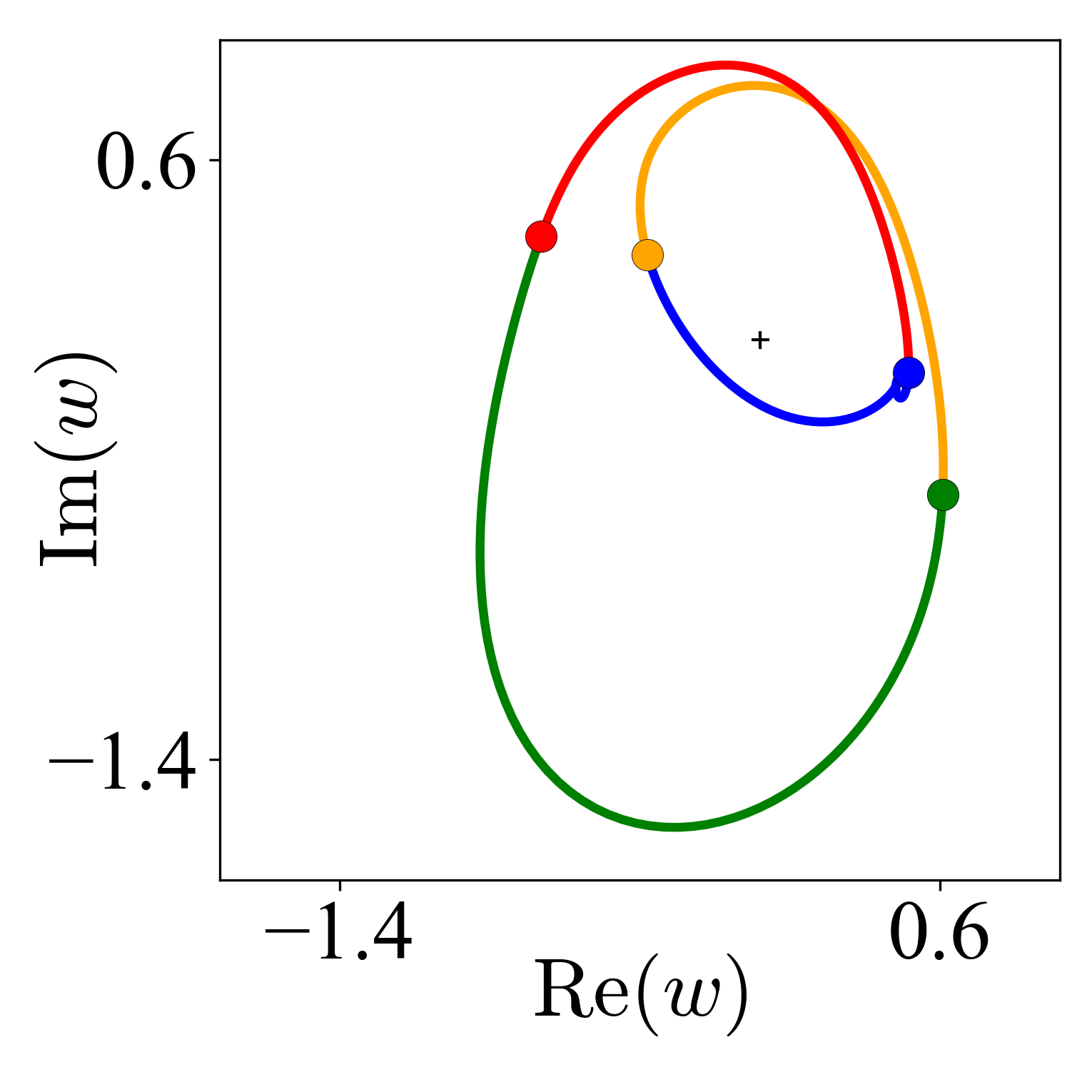}
    \hspace{-2.5mm}          
    \includegraphics[width=0.250\columnwidth]{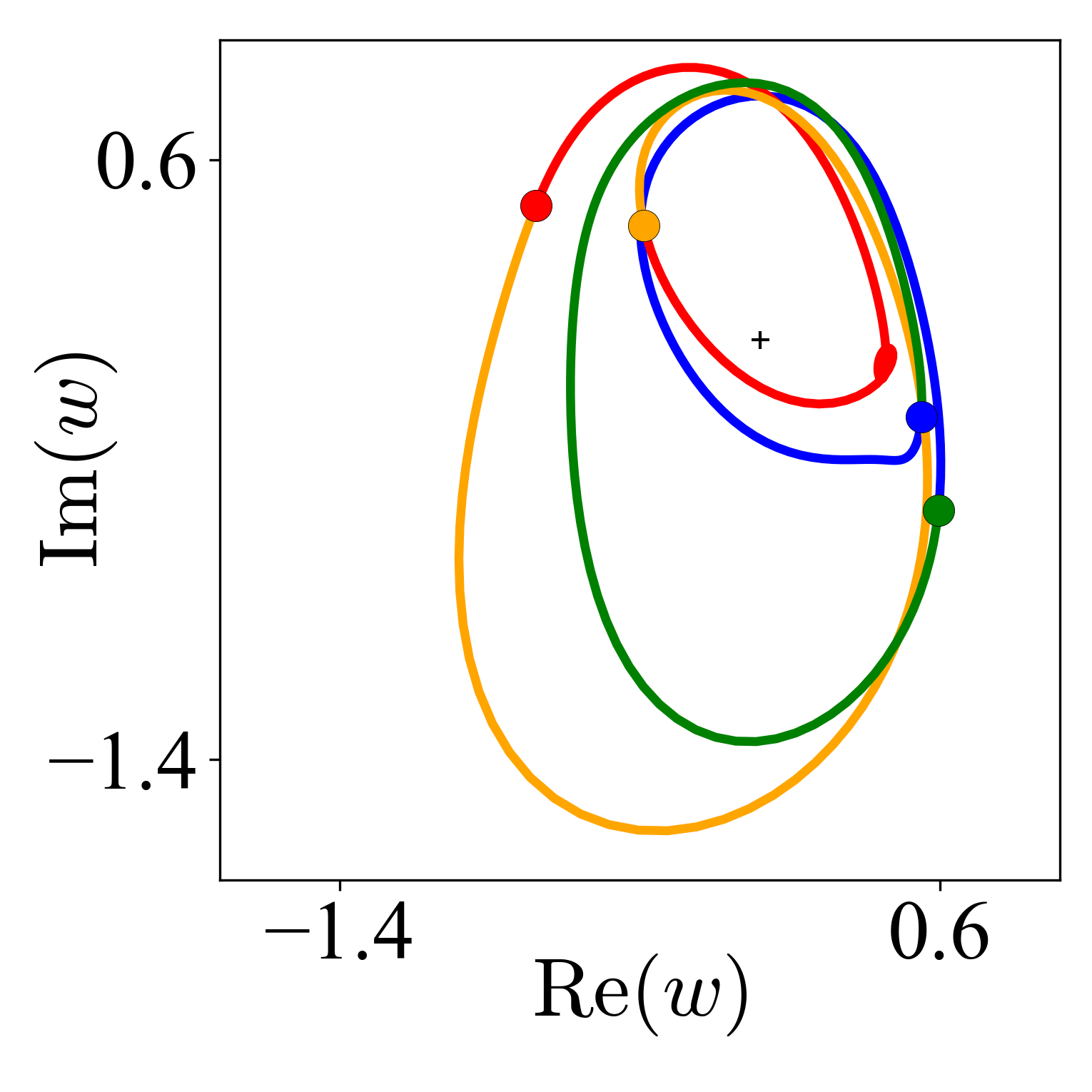}
    \hspace{-2.5mm}          
    \includegraphics[width=0.250\columnwidth]{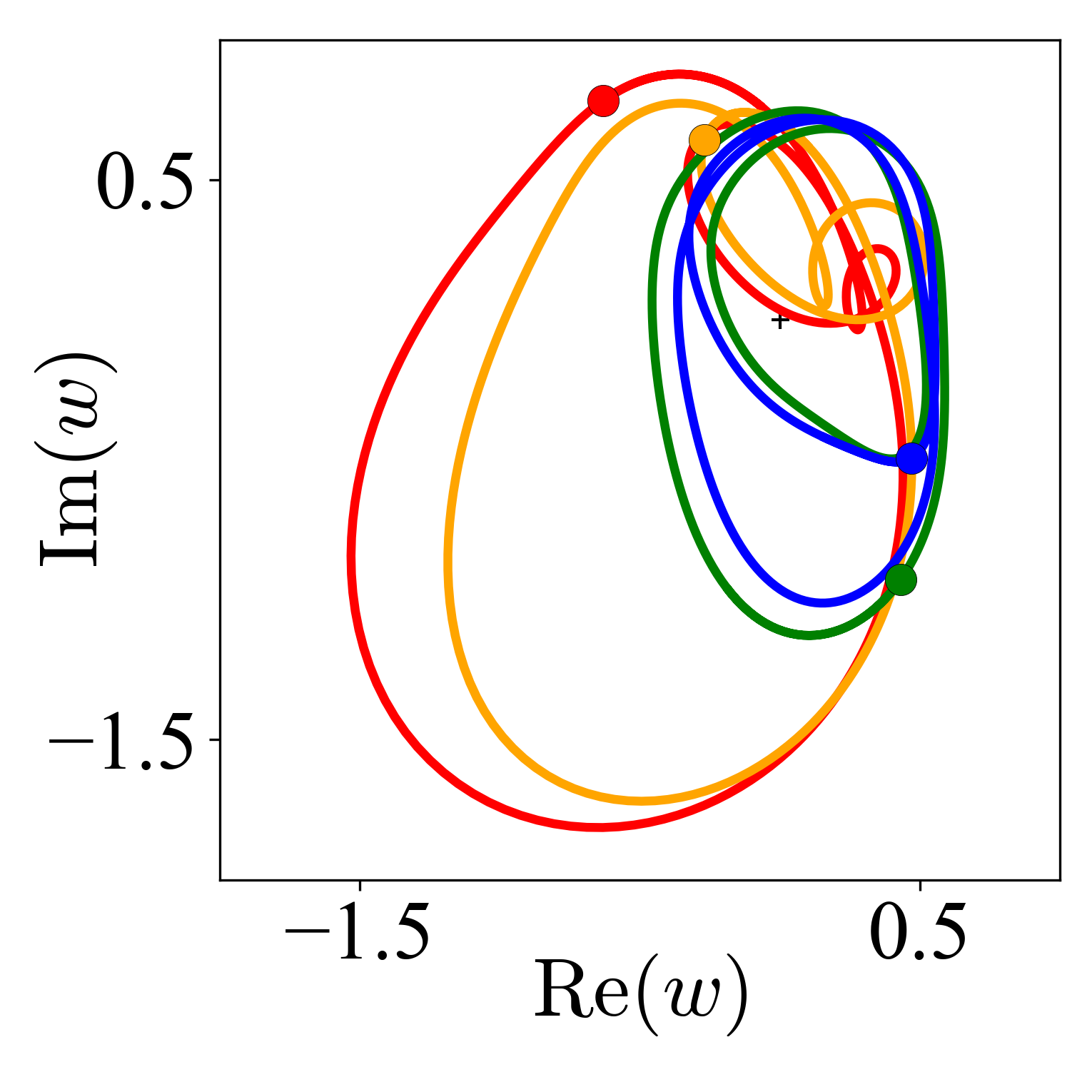}
    \hspace{-2.5mm}          
    \includegraphics[width=0.250\columnwidth]{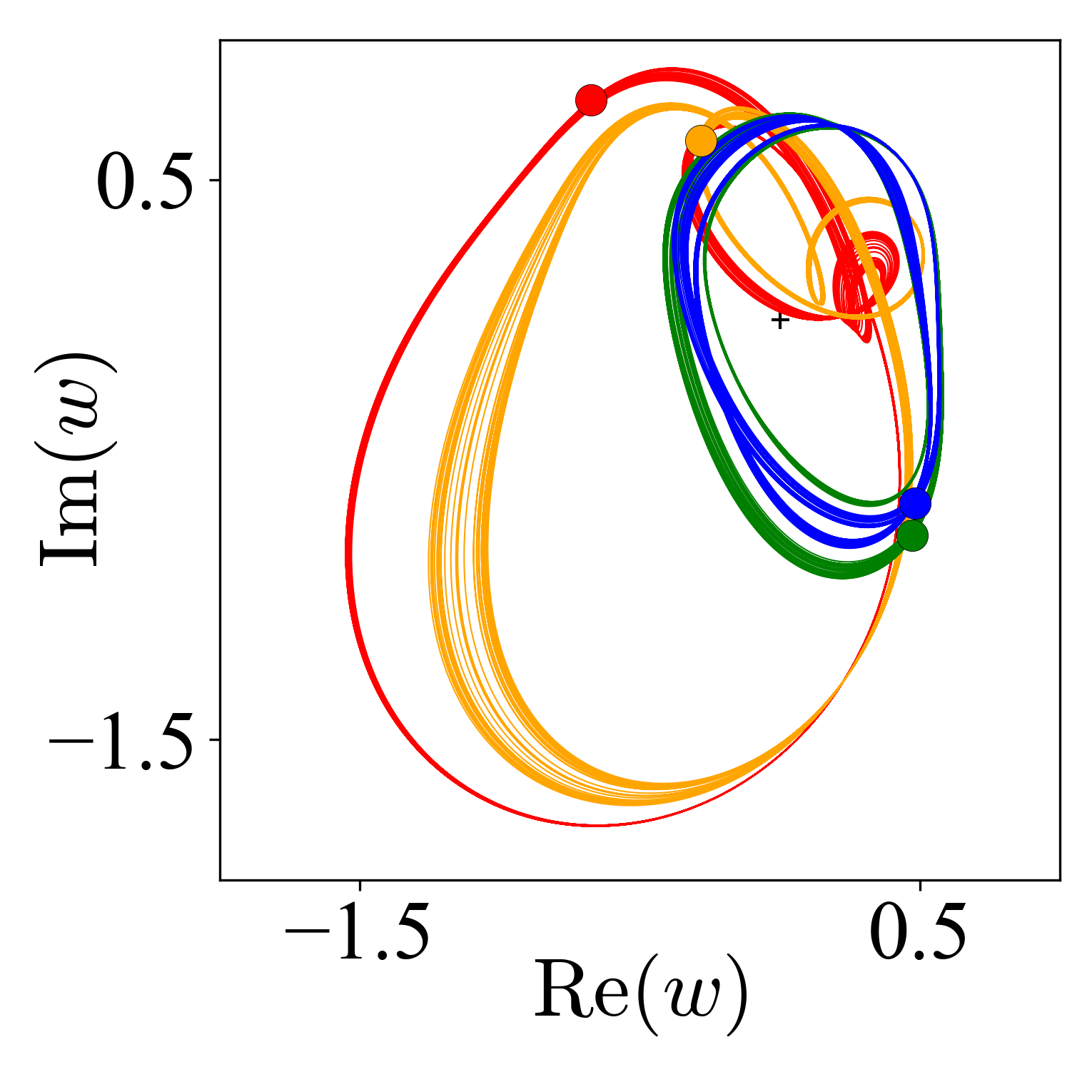}
    \vspace{-2.5mm}
    \end{minipage}
        };

        \node at (-6.70,  3.65) {$\mathbb{S}_2\times\mathbb{S}_2\rtimes\mathbb{Z}_2$};
        \node at (-2.80,  3.65) {$\mathbb{S}_2\times\mathbb{Z}_2$};
        \node at ( 1.10,  3.65) {$\mathbb{S}_2$};
        \node at ( 5.35,  3.65) {$\mathbb{S}_2$};
        \node at (-7.40, -0.56) {$\mathbb{Z}_4$};
        \node at (-3.15, -0.56) {$\mathbb{Z}_2$};
        \node at ( 1.10, -0.56) {\small $\{e\}$};
        \node at ( 5.35, -0.56) {\small $\{e\}$};

        \node[align=center, anchor=south] at (-8.16,  3.85) {\small (a)};
        \node[align=center, anchor=south] at (-3.90,  3.85) {\small (b)};
        \node[align=center, anchor=south] at ( 0.33,  3.85) {\small (c)};
        \node[align=center, anchor=south] at ( 4.55,  3.85) {\small (d)};
        \node[align=center, anchor=south] at (-8.16, -0.35) {\small (e)};
        \node[align=center, anchor=south] at (-3.90, -0.35) {\small (f)};
        \node[align=center, anchor=south] at ( 0.33, -0.35) {\small (g)};
        \node[align=center, anchor=south] at ( 4.55, -0.35) {\small (h)};

    \end{tikzpicture}
    \caption{
    Complex-plane trajectories of different $N=4$ solutions.
    Each color corresponds to an oscillator, lines to trajectories and filled circles to %
    current
    positions.
    Labels in the upper right denote %
    interchange symmetries of the oscillators.
    All solutions are viewed in the co-rotating frame of the ensemble average $\langle W \rangle$.
    (a)~$2\!-\!2$ period-1 solution for $c_2=-0.78$ and $\eta=0.62$.
    (b)~$2\!-\!1\!-\!1$ period-2 solution emerging from (a). 
    $c_2=-0.7$ and $\eta=0.62$.
    (c)~Less symmetric $2\!-\!1\!-\!1$ period-2 solution emerging from (b). 
    $c_2=-0.64$ and $\eta=0.615$.
    (d)~Chaotic $2\!-\!1\!-\!1$ solution for $c_2=-0.626$ and $\eta=0.6185$.
    (e)~$1\!-\!1\!-\!1\!-\!1$ rotating-wave period-2 solution emerging from (a). 
    $c_2=-0.7$ and $\eta=0.6$.
    (f)~Less symmetric $1\!-\!1\!-\!1\!-\!1$ period-2 solution mediating between (b) and (e). 
    $c_2=-0.7$ and $\eta=0.61$.
    (g)~Fully asymmetric period-2 solution mediating between (c) and (f). 
    $c_2=-0.63$ and $\eta=0.615$.
    (h)~Fully asymmetric chaotic solution for $c_2=-0.626$ and $\eta=0.618$.
    $\nu=0.1$ throughout.
    }
    \label{fig:N4_CP_plots}
\end{figure*}
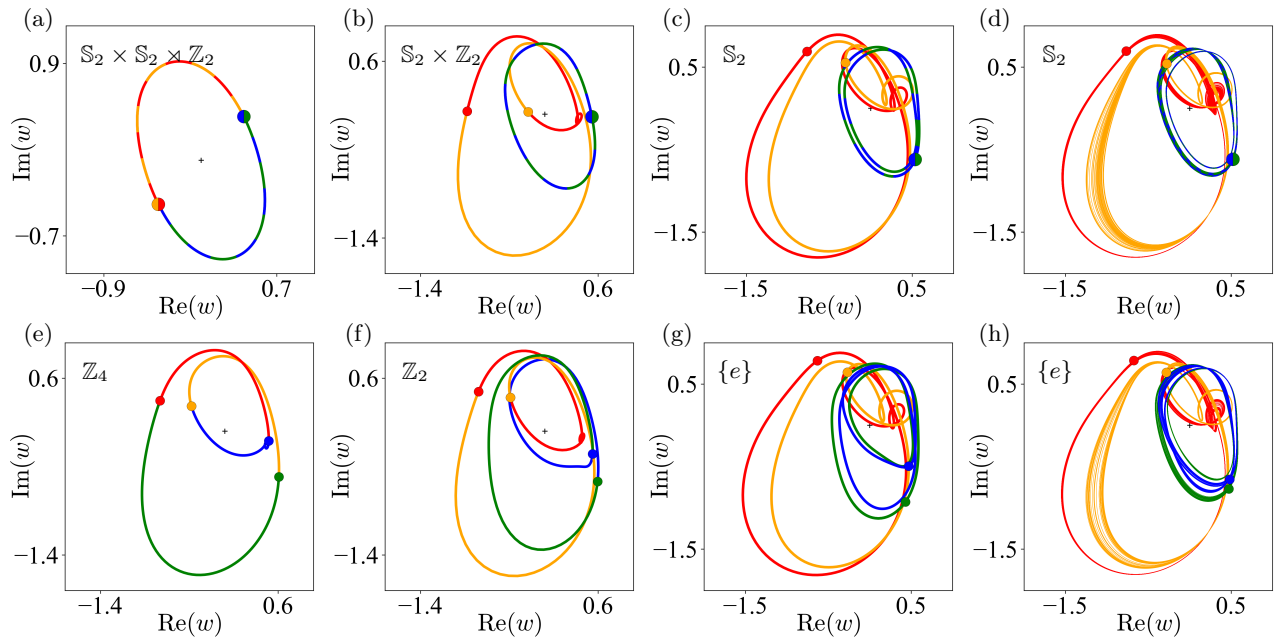

Our model is an ensemble of $N$ Stuart-Landau oscillators $W_k\in\mathbb{C}$, $k=1,\dots,N$,
with nonlinear global coupling~\cite{Schmidt_Chaos_2014}:
\begin{equation}\label{eq:SLE}
\begin{aligned}
    \frac{\mathrm{d}W_k}{\mathrm{d}t}~=~
    &W_k - (1 + i c_2)|W_k|^2W_k \\
    &- (1 + i \nu)\langle W \rangle + (1 + i c_2)\langle |W|^2W \rangle\,,
\end{aligned}
\end{equation}
\noindent where $\langle\dots\rangle = 1/N\sum_{k=1}^N \dots$ denotes ensemble averages while $c_2$ and $\nu$ are real parameters. %
The Stuart-Landau oscillator itself %
is a generic %
model for a system
close to a Hopf bifurcation, that is, to the onset of self-sustained oscillations~\cite{Kuramoto_Book_1984}.
Because the oscillators are identical and only affect each other through the mean quantities $\langle W \rangle$ and $\langle |W|^2 W \rangle$, Eq.~\eqref{eq:SLE} is %
$\mathbb{S}_N$-equivariant:
If $\mathbf{W}(t)\in\mathbb{C}^N$ is a solution, %
then so is $\gamma\mathbf{W}(t)~\forall\,\gamma\in\mathbb{S}_N$, where $\mathbb{S}_N$, denoted the \emph{symmetric group}, is the group of all permutations of the $N$ oscillators~\cite{Moehlis_ScholPed_2007}. 
When taking the ensemble average of the equations~\eqref{eq:SLE},
we further find that 
\begin{equation}\label{eq:average_motion}
    \langle \frac{\mathrm{d}W_k}{\mathrm{d}t} \rangle = \frac{\mathrm{d}}{\mathrm{d}t}\langle W \rangle = - i \nu \langle W \rangle 
    ~~\Rightarrow~~ \langle W \rangle = \eta\, \mathrm{e}^{-i \nu t}\,,
\end{equation}
\noindent which implies that $\langle W \rangle$ is
confined to simple harmonic motion with frequency $\nu$.
Its amplitude $\eta\in\mathbb{R}$ is implicitly set by choosing the initial condition and thus constitutes an additional parameter.
Eq.~\eqref{eq:average_motion} also implies that there always exists a fully synchronized periodic solution $W_k = \langle W \rangle = \eta \mathrm{e}^{-i \nu t}~\forall\, k$.
For $N=1$, this is clearly the only solution possible, while for $N>2$, it has been shown to lose stability and give rise to different two-cluster solutions~\cite{Schmidt_PRE_2014}:
This happens either in an equivariant pitchfork bifurcation, producing separate clusters that continue to circle the origin with frequency $\nu$ at different fixed amplitudes,
or in an equivariant secondary Hopf bifurcation, producing separate modulated-amplitude clusters that henceforth oscillate with two superposed frequencies $\nu$ and $\omega_\mathrm{H}$.
For large $N$, each of these cluster states has previously been deduced to somehow produce its own kind of chimera states, labeled type-I and type-II chimeras, respectively~\cite{Schmidt_PRL_2015}.
Most recently, the concrete path from modulated-amplitude two-cluster states to macroscopic type-II chimeras was uncovered~\cite{Haugland_NatComm_2021}.

Here, we restrict ourselves to the case $N=4$, and 
again focus on modulated-amplitude dynamics with two component oscillations.
We also keep $\nu=0.1$ fixed.
Because the ensemble average $\langle W \rangle = \eta \mathrm{e}^{-i \nu t}$ is independent of the individual oscillator dynamics, it describes an always accessible rotating frame of reference, wherein the value of each oscillator is always given by the following relation: 
\begin{equation}\label{eq:transfer_to_rotating}
    W_k = \eta\,\mathrm{e}^{-i \nu t}(1 + w_k) ~~ \Rightarrow ~~ w_k = W_k \eta^{-1}\mathrm{e}^{i \nu t} - 1\,,
\end{equation}
\noindent where $w_k$ is the value of the oscillator $W_k$ in the co-rotating frame %
and $\sum_k w_k = 0$.
Viewing the dynamics in the %
frame of $\langle W \rangle$ %
means that 
two-frequency
quasiperiodic motion %
becomes a simple limit cycle.
A $2\!-\!2$ originally modulated-amplitude solution with two oscillators in each of two clusters thus looks as in Fig.~\ref{fig:N4_CP_plots}\,a.
The cross %
denotes the position of the ensemble average, which in the co-rotating frame %
is always at the origin.
Because the two clusters of the $2\!-\!2$ solution are of equal size, their respective displacement from %
the origin
is always equal and opposite.
Furthermore, they follow the same %
trajectory with a mutual phase shift of half their common~period.

The symmetry (or isotropy subgroup~\cite{Hoyle_Book_2006}%
) of the $2\!-\!2$ solution in the rotating frame is $\mathbb{S}_2\times\mathbb{S}_2\rtimes\mathbb{Z}_2$, where $\mathbb{S}_2$ denotes the permutations of the oscillators within either cluster, operations that do not change the current solution.
If we interchange the clusters themselves, this does change the instantaneous positions of the individual oscillators.
Yet, it projects each oscillator exactly onto the the point where it would normally be half a period later. 
Combining this permutation with a time shift of half a period thus also leaves the system state unchanged.
We denote this kind of spatiotemporal component symmetry by $\mathbb{Z}_2$, the symmetry $\mathbb{Z}_n$ of a cyclic group of $n=2$ elements.

The permutation symmetries $\mathbb{S}_2$ %
combine %
by means of a direct product ``$\times$'' because their operations commute.
In contrast, interchanging the oscillators of the first cluster and then interchanging the clusters is not the same operation as first interchanging the clusters and then interchanging the oscillators of the first cluster.
(Of course, the two-cluster solution in Fig.~\ref{fig:N4_CP_plots}\,a remains the same in either case, but the mapping from old to new oscillator indices does not.)
Therefore, %
$\mathbb{S}_2 \times \mathbb{S}_2$ and $\mathbb{Z}_2$
combine
with a so-called semidirect product ''$\rtimes$``~\cite{Moehlis_ScholPed_2007}.
In total,
the $2\!-\!2$ solution in Fig.~\ref{fig:N4_CP_plots}\,a is invariant under %
\begin{equation}
    |\mathbb{S}_2|\cdot|\mathbb{S}_2|\cdot|\mathbb{Z}_2| = 2\cdot2\cdot2 = 8
\end{equation}
different symmetry operations (including the identity),
where $|\Gamma|$ denotes the number of elements in the group $\Gamma$.
The fully synchronized solution $w_k=0~\forall\,k$ is invariant under a total of $|\mathbb{S}_4| = 4!%
= 24$ permutations. 
Those permutations in $\mathbb{S}_4$ that are not contained in $\mathbb{S}_2\times\mathbb{S}_2\rtimes\mathbb{Z}_2$ transform the current $2\!-\!2$ solution into %
one
out of 
two 
distinct, but equivalent solution variants,
whose overall dynamics are %
the same as
those in Fig.~\ref{fig:N4_CP_plots}\,a, but wherein not the yellow, but either the blue or the green oscillator are clustered with the red one.
All equivalent solution variants together form the so-called group orbit ~\cite{Hoyle_Book_2006}.
As the equations themselves are $\mathbb{S}_4$-equivariant, the size of any solution's isotropy subgroup multiplied by the size of its group orbit is always $|\mathbb{S}_4|=24$.

If $c_2$ is sufficiently increased for %
suitable values of %
$\eta$,
the $2\!-\!2$ solution loses stability in an equivariant 
period-doubling bifurcation.
For %
$\eta \in [0.62, 0.7]$, this produces a stable $2\!-\!1\!-\!1$ solution like the one in Fig.~\ref{fig:N4_CP_plots}\,b.
Thereby, either of the clusters of two is split, and the resultant single oscillators henceforth undergo the same kind of period-2 motion, 
with double the period of the remaining intact cluster.
The mutual phase shift of the single oscillators is exactly half their common period, ensuring the overall symmetry $\mathbb{S}_2\times\mathbb{Z}_2$.
As either of the two clusters %
can be
the one to split up, 
two $\mathbb{S}_2\times\mathbb{Z}_2$ variants emerge from each variant of the $2\!-\!2$ solution.

For appropriate values of $\eta$,
additionally increasing $c_2$ breaks the $\mathbb{Z}_2$ symmetry by differentiating the trajectories of the single oscillators.
Because of the constraint $\sum_k w_k = 0$, this also means that the trajectory of the intact cluster becomes period-2, 
as shown in Fig.~\ref{fig:N4_CP_plots}\,c,
but the overall periodicity of the ensemble remains the same.
As either of the %
single oscillators %
can be
the one to start traveling along the largest loop in the complex plane, this change is an equivariant pitchfork bifurcation,
with $|\mathbb{S}_2\times\mathbb{Z}_2|/|\mathbb{S}_2|=2$ equivalent $\mathbb{S}_2$ variants emerging from each $\mathbb{S}_2\times\mathbb{Z}_2$ variant.

Notably, the $2\!-\!1\!-\!1$ solution in Fig.~\ref{fig:N4_CP_plots}\,b is not the only outcome of %
the equivariant period-doubling of the $2\!-\!2$ solution.
Below $\eta=0.62$, increasing $c_2$ instead causes the break-up of both clusters, and a stable $1\!-\!1\!-\!1\!-\!1$ solution is %
produced.
Here, the four single oscillators all proceed along the same trajectory, with consecutive phase shifts of a quarter %
period, as shown in Fig.~\ref{fig:N4_CP_plots}\,e.
The symmetry is thus a cyclic %
$\mathbb{Z}_4$ symmetry.

Comparing the $\mathbb{S}_2\times\mathbb{Z}_2$ and the $\mathbb{Z}_4$ solution, we realize that the former might be turned into the latter if its remaining cluster were to split and the resultant single oscillators to become period-2. 
If $\eta$ is decreased from the region where the $\mathbb{S}_2\times\mathbb{Z}_2$ solution is stable, such a split does indeed occur, and the %
solution in Fig.~\ref{fig:N4_CP_plots}\,f is produced.
It symmetry is $\mathbb{Z}_2$ because only the operation of exchanging both the blue and the green, and the red and the yellow oscillator, respectively, followed by a time shift of half %
a
period, leaves the current solution unchanged.
As $\eta$ is further decreased, the less pronouncedly period-2 trajectory of the blue and the green oscillator becomes ever more similar to that of the red and the yellow oscillator, until we reach the %
$\mathbb{Z}_4$ solution.
Supplementary Fig. 1 details which $\mathbb{Z}_2$ variants mediate between the two $\mathbb{S}_2\times\mathbb{Z}_2$ and two $\mathbb{Z}_4$ variants emerging from the same $2\!-\!2$ variant.

Similarly, the $\mathbb{S}_2$ solution in Fig.~\ref{fig:N4_CP_plots}\,c may also lose its remaining symmetry in an equivariant pitchfork affecting the %
remaining
cluster.
This results in the fully asymmetric solution in Fig.~\ref{fig:N4_CP_plots}\,g.
From here on, the loops of the blue and the green, and the red and the yellow oscillator, respectively, may also become more similar again, and for suitable values of $c_2$ and $\eta$, the fully asymmetric solution %
becomes
a $\mathbb{Z}_2$ solution like that in Fig.~\ref{fig:N4_CP_plots}\,f.
How different fully asymmetric variants mediate between the different $\mathbb{S}_2$ and $\mathbb{Z}_2$ variants is shown in Supplementary Fig.~2.

\begin{figure*}[ht]
    \centering
    \vspace{-1.0mm}
    \begin{tikzpicture}
        \node (fig1) at (0,0)
        {\centering
    \begin{minipage}{2.035\columnwidth}
    \centering
    \hspace{-1mm}
    \includegraphics[width=0.49\textwidth]{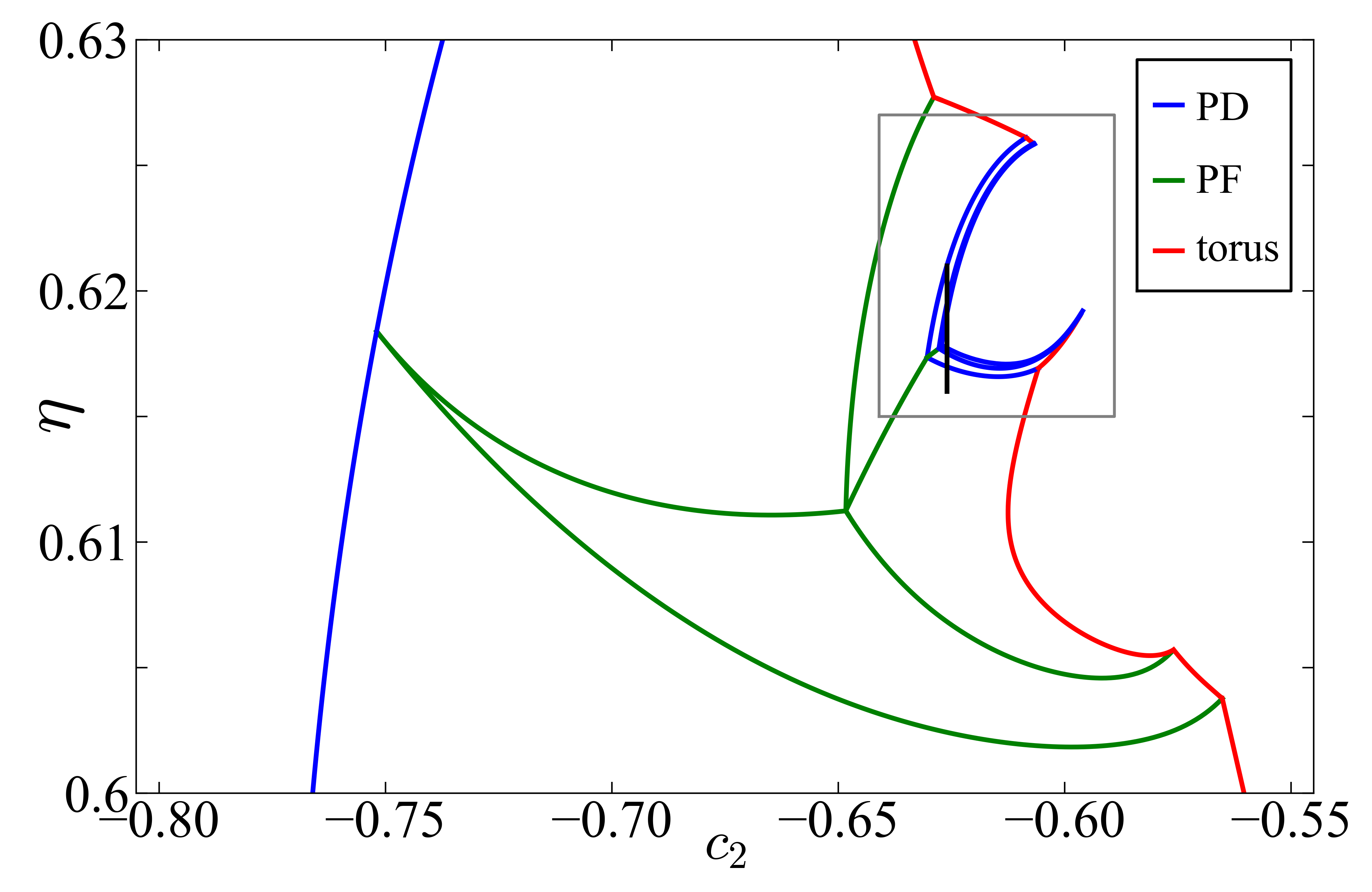}
    \hspace{-3.0mm}
    \includegraphics[width=0.49\textwidth]{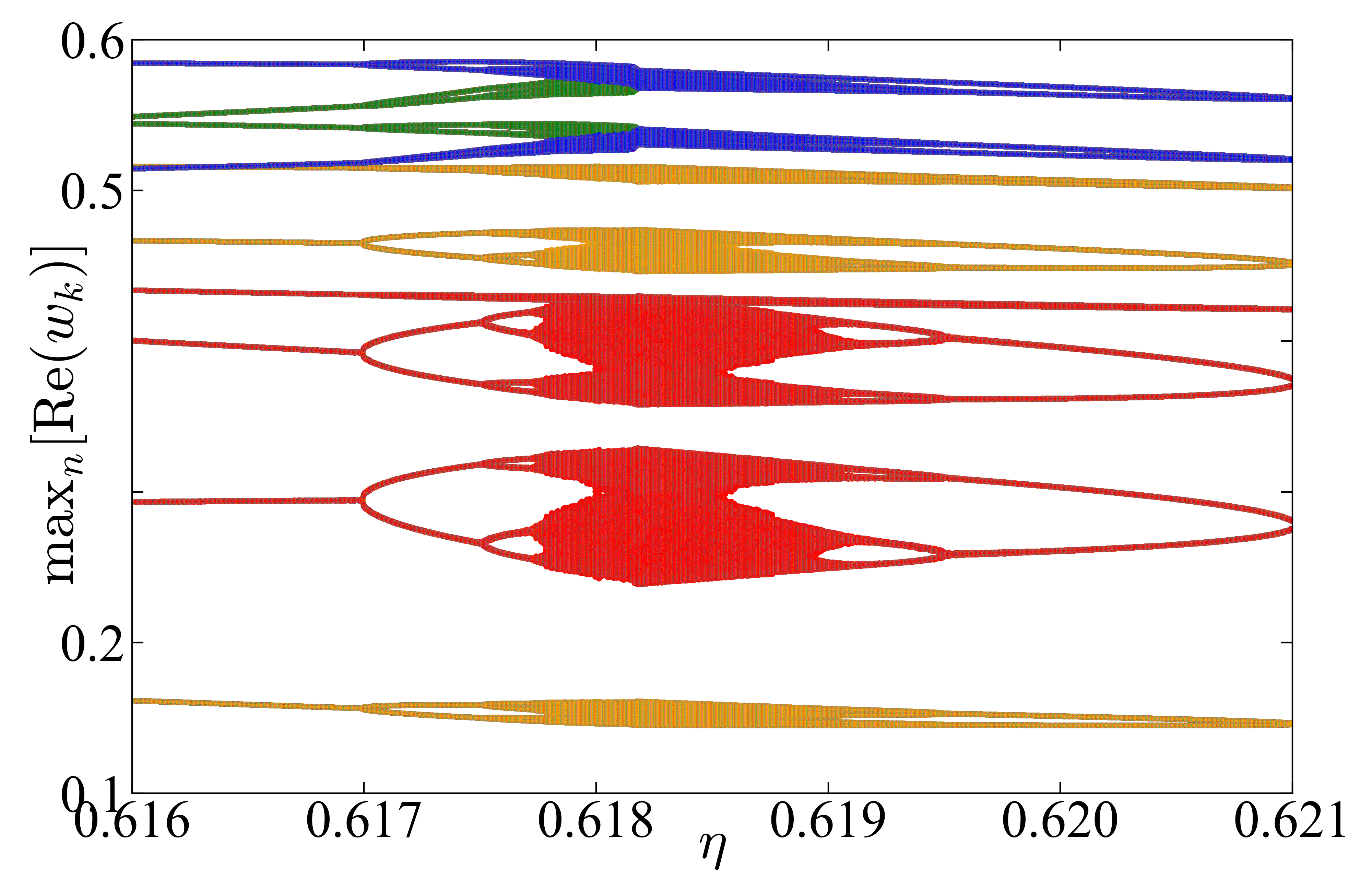}\\
    \vspace{02mm}
    \hspace{-1mm}
    \includegraphics[width=0.49\textwidth]{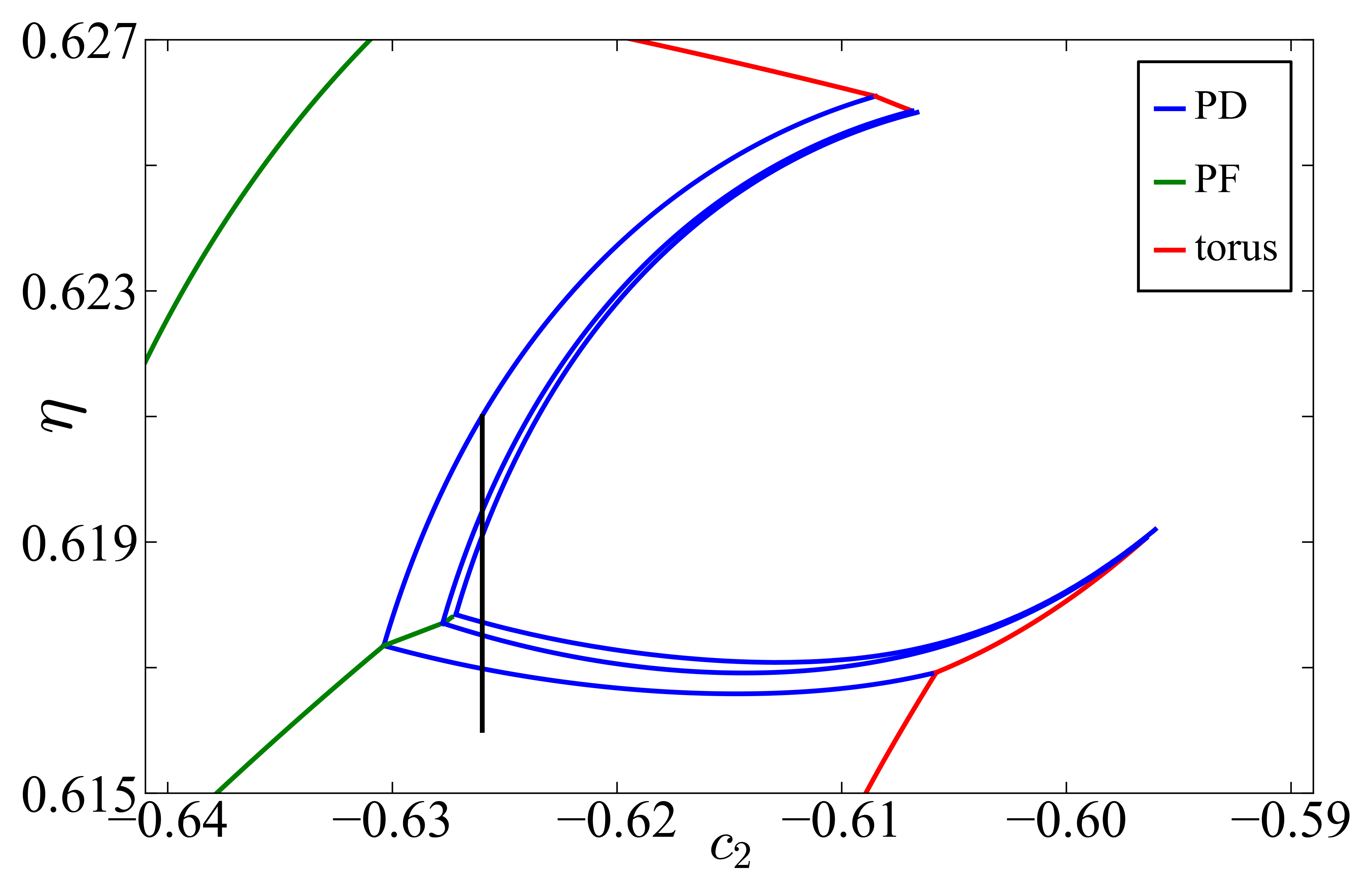}
    \hspace{-2.3mm}
    \includegraphics[width=0.485\textwidth]{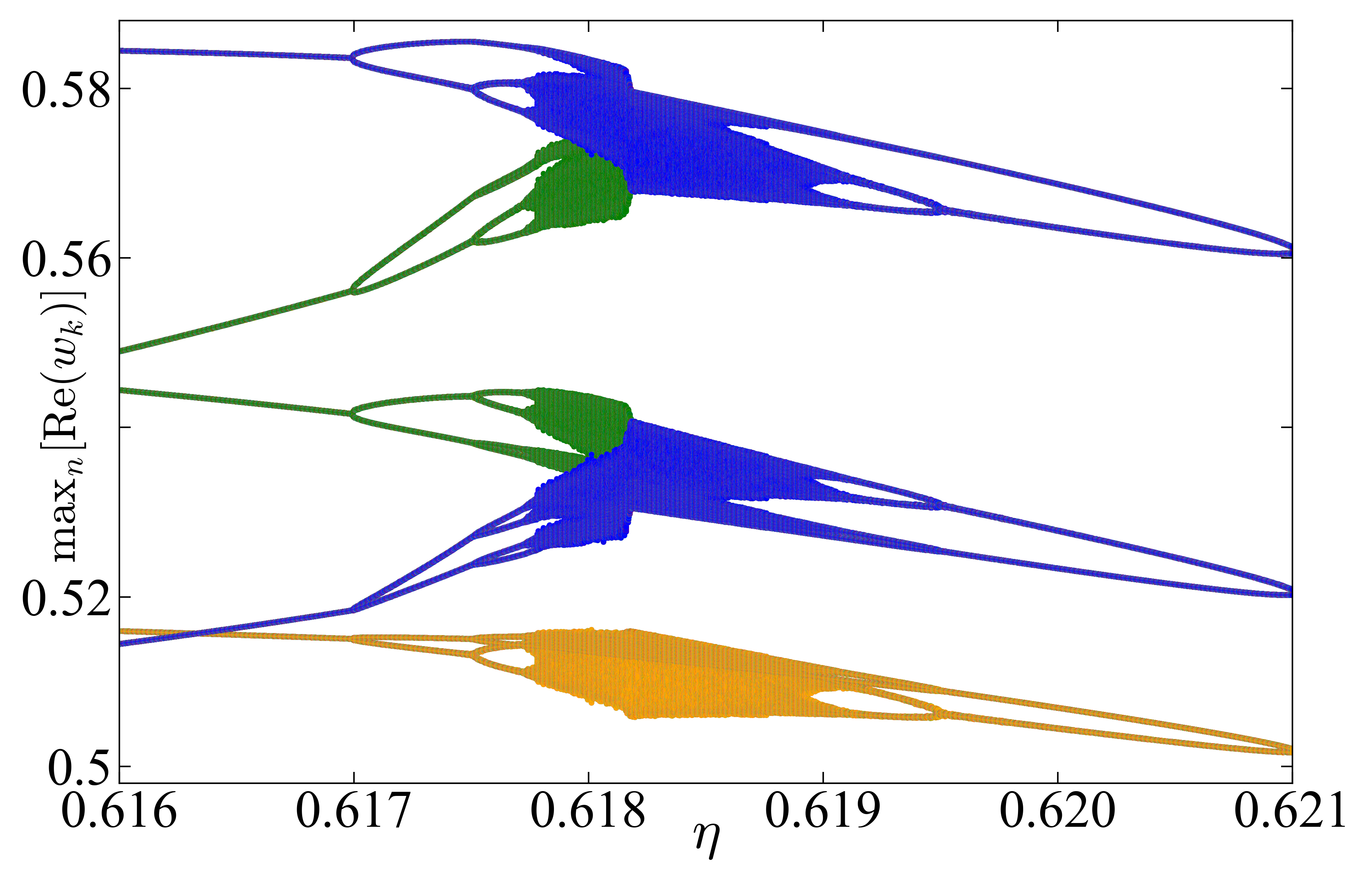}
    \end{minipage}
    \vspace{-5mm}
        };

        \node[align=center, anchor=center] at (-6.9,  4.80) {\large $\mathbb{S}_2\times\mathbb{S}_2$\\
                                                            \large $\rtimes\mathbb{Z}_2$};
        \node at (-4.60,  4.00) {\large $\mathbb{S}_2\times\mathbb{Z}_2$};
        \node at (-5.60,  1.50) {\large $\mathbb{Z}_4$};
        \node at (-3.60,  2.00) {\large $\mathbb{Z}_2$};
        \node at (-2.75,  4.50) {$\mathbb{S}_2$};
        \node at (-2.70,  2.60) {\large $\{e\}$};

        \node at (-6.30, -2.00) {\Large $\mathbb{S}_2$};
        \node at (-5.00, -4.80) {\large $\{e\}$};

        \node[align=center, anchor=south] at (-7.45,  5.50) {(a)};
        \node[align=center, anchor=south] at ( 1.05,  5.50) {(b)};
        \node[align=center, anchor=south] at (-7.45, -0.35) {(c)};
        \node[align=center, anchor=south] at ( 1.05, -0.35) {(d)};
        
    \end{tikzpicture}
    \caption{
    (a)~Bifurcation diagram showing the regions of stability of the solutions in Fig.~\ref{fig:N4_CP_plots} and the bifurcations between them.
    PD = period-doubling.
    PF = pitchfork.
    $\nu=0.1$.
    (b)~Maxima of $\mathrm{Re}(w_k)$ for each oscillator $k=1,\dots,4$ as $\eta$ is gradually increased along the vertical black line at $c_2=-0.626$ in (a).
    At $\eta=0.6182$, the blue and green oscillator merge in a pitchfork bifurcation of chaotic attractors.
    (c)~Magnification of the region within the gray rectangle in (a).
    (d)~Upper part of (b) showing the increase in symmetry in greater detail.
    The bifurcation lines in (a) and (c) were calculated using Auto07p~\cite{Doedel_Report_2012}.
    }
    \label{fig:N4_2+2_bifurcations}
\end{figure*}
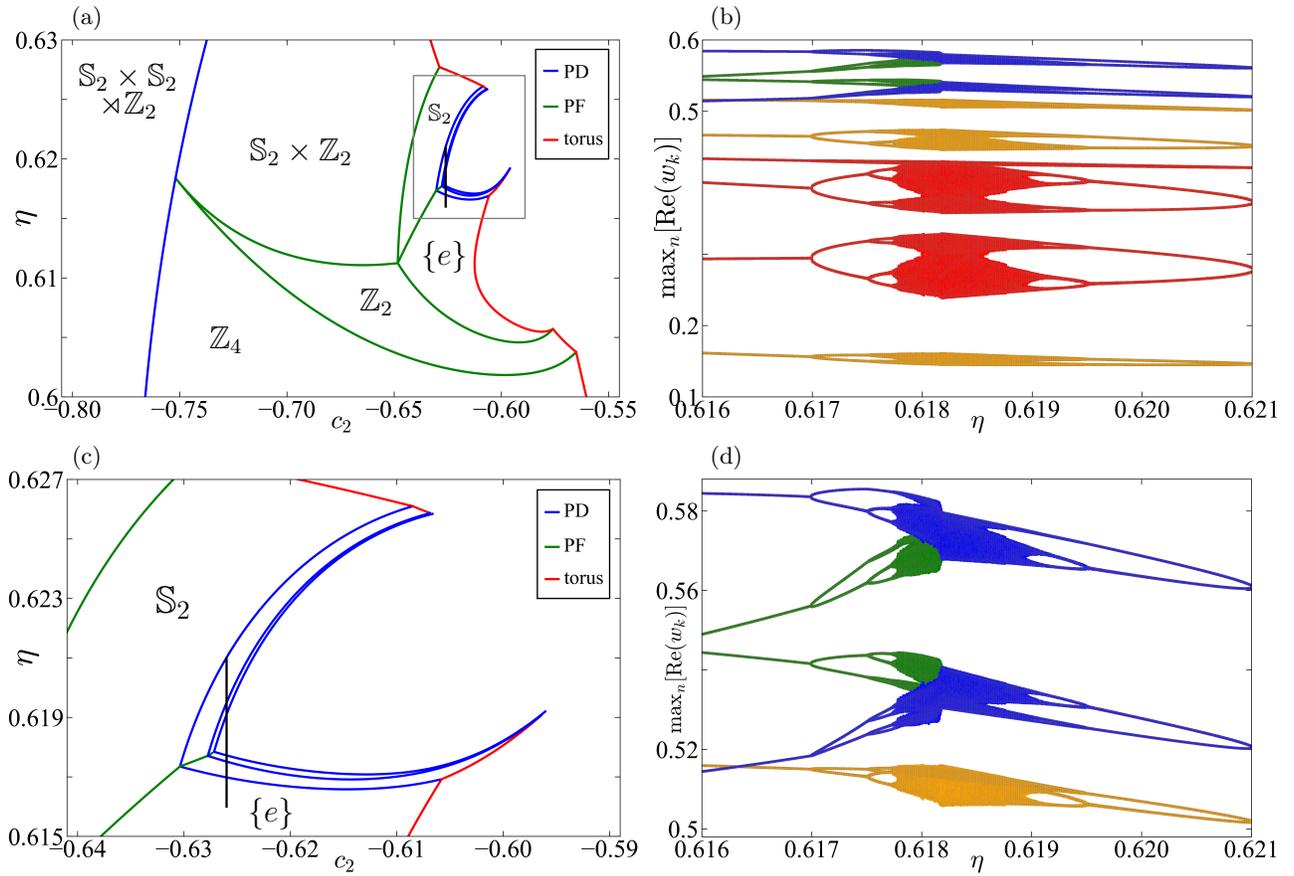

In parameter space, the stable regions of these
solutions %
are connected 
as shown in Fig.~\ref{fig:N4_2+2_bifurcations}\,a, 
where the equivariant period-doubling of the $2\!-\!2$ solution is denoted by the leftmost blue line and the equivariant pitchfork bifurcations between the different period-2 solutions by the green lines. 
Additional features of this bifurcation diagram are the torus bifurcations shown in red as well as the collection of blue period-doublings in the upper right.
Each of the former simply adds a tertiary frequency to the adjoining period-2 solution, causing the system to undergo $T^3$ quasiperiodic motion overall, that is, two-frequency quasiperiodic motion in the frame of $\langle W \rangle$.
A little further to the right, this motion also becomes unstable and the ensemble jumps to a multistable $3\!-\!1$ solution.

The parameter region of the extra period-doublings is shown in greater detail in Fig.~\ref{fig:N4_2+2_bifurcations}\,c.
Here, it becomes apparent that these are really two adjacent period-doubling cascades.
The starting point for the upper cascade is the $\mathbb{S}_2$ solution in Fig.~\ref{fig:N4_CP_plots}\,c while that of the lower one is the fully asymmetric solution in Fig.~\ref{fig:N4_CP_plots}\,g.
Unlike the period-doubling in the left of Fig.~\ref{fig:N4_2+2_bifurcations}\,a, these period-doubling bifurcations do not affect the symmetry of the solution, but simply double the period of every oscillator.
Thus, if we start at the $\mathbb{S}_2$ period-2 solution and increase $c_2$ or decrease $\eta$ appropriately, we first obtain an $\mathbb{S}_2$ period-4 solution, then an $\mathbb{S}_2$ period-8 solution, and so on, until we finally end up at the chaotic $\mathbb{S}_2$ solution in Fig.~\ref{fig:N4_CP_plots}\,d.
This is a minimal chimera state for which the underlying chaotic attractor is not symmetric under the exchange of the two unsynchronized oscillators~\cite{Chossat_PhysicaD_1988}. %
If we start at the asymmetric period-2 solution in Fig.~\ref{fig:N4_CP_plots}\,g and increase $\eta$, the system undergoes a period-doubling cascade to a chaotic state in which the trajectories of all four oscillators differ.
See Fig.~\ref{fig:N4_CP_plots}\,h.

Let us now consider the region of the bifurcation diagram in
Fig.~\ref{fig:N4_2+2_bifurcations}\,c where the two period-doubling cascades meet.
Above, we saw that %
each variant of the $\mathbb{S}_2$ period-2 solution bifurcates in an equivariant pitchfork to two distinct, but equivalent variants of the fully asymmetric period-2 solution.
Similarly, the region where $\mathbb{S}_2$ period-4 solutions are stable borders on the region where fully asymmetric period-4 solutions are.
Every $\mathbb{S}_2$ period-$2^n$ solution %
also undergoes an equivariant pitchfork bifurcation, 
and in these equivariant pitchforks, two equivalent variants of the asymmetric period-$2^n$ solutions emerge from each variant of the $\mathbb{S}_2$-symmetric period-$2^n$ solution.

Continuing along the juncture of the two period-doubling cascades, the succession of equivariant pitchforks finally reaches the chaotic domain.
If initializing the ensemble in the asymmetric period-2 solution at $c_2=-0.626$ and $\eta=0.616$ and slowly increasing $\eta$ along the vertical black line in Fig.~\ref{fig:N4_2+2_bifurcations}\,c, we thus observe the development depicted in Fig.~\ref{fig:N4_2+2_bifurcations}\,b and~d.
Here, each oscillator is represented by the maxima reached by its real part $\mathrm{Re}(w_k)$.
Initially, each oscillator reaches either two or three distinct maxima. %
At $\eta=0.617$, the number of these maxima are doubled,
at $\eta=0.6175$, they are doubled again,
and at $\eta=0.6177$ again.
More distinct period-doublings cannot be discerned, but the overall resemblance to the classic Feigenbaum period-doubling cascade remains clear.

The crucial point of the development happens at $\eta=0.6182$, where the current fully asymmetric chaotic solution merges with one of its own variants -- here: the one in which the blue and green oscillator are interchanged -- to form a single variant of a $2\!-\!1\!-\!1$ chaotic chimera state.
Continuing further upward in $\eta$ takes us 
full circle 
from this chimera backwards through the $\mathbb{S}_2$ period-doubling cascade to the $\mathbb{S}_2$ period-2 solution in Fig.~\ref{fig:N4_CP_plots}\,c 
and the other regular solutions. %

In summary, we have uncovered the interconnected bifurcation structure of differently symmetric and differently periodic solutions arising in an $\mathbb{S}_4$-symmetric system.
Embedded in this structure are minimal %
chaotic chimera states.
The bifurcation-theoretical route %
to these minimal chimera states differs from that of the macroscopic chimeras found in the same model for larger ensemble sizes $N$~\cite{Haugland_NatComm_2021}, a trait they incidentally share with the otherwise unrelated minimal chimeras found in the two-groups model of phase oscillators~\cite{Abrams_PRL_2008,Panaggio_PRE_2016}.
Our minimal chimeras strongly resemble the so-called asymmetric chimera states previously reported in Stuart-Landau oscillators with linear global coupling~\cite{Kemeth_PRL_2018},
whose context of other solutions is not as extensively charted.
Notably, our %
bifurcation structure contains a juncture of two period-doubling cascades, leading to equivariant pitchfork bifurcations of chaotic attractors and rendering the chimera states reachable by means of two different routes in parameter space.
To what extent this is a universal property of $\mathbb{S}_4$-symmetric systems remains an interesting question. 

The authors thank Felix P. Kemeth for fruitful discussions.
Financial support from the Studienstiftung des deutschen Volkes and the Deutsche Forschungsgemeinschaft, project KR1189/18 “Chimera States and Beyond”,
is gratefully acknowledged.

\bibliography{references_manually_edited.bib}

\begin{thebibliography}{30}%
\makeatletter
\providecommand \@ifxundefined [1]{%
 \@ifx{#1\undefined}
}%
\providecommand \@ifnum [1]{%
 \ifnum #1\expandafter \@firstoftwo
 \else \expandafter \@secondoftwo
 \fi
}%
\providecommand \@ifx [1]{%
 \ifx #1\expandafter \@firstoftwo
 \else \expandafter \@secondoftwo
 \fi
}%
\providecommand \natexlab [1]{#1}%
\providecommand \enquote  [1]{``#1''}%
\providecommand \bibnamefont  [1]{#1}%
\providecommand \bibfnamefont [1]{#1}%
\providecommand \citenamefont [1]{#1}%
\providecommand \href@noop [0]{\@secondoftwo}%
\providecommand \href [0]{\begingroup \@sanitize@url \@href}%
\providecommand \@href[1]{\@@startlink{#1}\@@href}%
\providecommand \@@href[1]{\endgroup#1\@@endlink}%
\providecommand \@sanitize@url [0]{\catcode `\\12\catcode `\$12\catcode
  `\&12\catcode `\#12\catcode `\^12\catcode `\_12\catcode `\%12\relax}%
\providecommand \@@startlink[1]{}%
\providecommand \@@endlink[0]{}%
\providecommand \url  [0]{\begingroup\@sanitize@url \@url }%
\providecommand \@url [1]{\endgroup\@href {#1}{\urlprefix }}%
\providecommand \urlprefix  [0]{URL }%
\providecommand \Eprint [0]{\href }%
\providecommand \doibase [0]{https://doi.org/}%
\providecommand \selectlanguage [0]{\@gobble}%
\providecommand \bibinfo  [0]{\@secondoftwo}%
\providecommand \bibfield  [0]{\@secondoftwo}%
\providecommand \translation [1]{[#1]}%
\providecommand \BibitemOpen [0]{}%
\providecommand \bibitemStop [0]{}%
\providecommand \bibitemNoStop [0]{.\EOS\space}%
\providecommand \EOS [0]{\spacefactor3000\relax}%
\providecommand \BibitemShut  [1]{\csname bibitem#1\endcsname}%
\let\auto@bib@innerbib\@empty
%</preamble>
\bibitem [{\citenamefont {Kaneko}(1990)}]{Kaneko_PhysicaD_1990}%
  \BibitemOpen
  \bibfield  {author} {\bibinfo {author} {\bibfnamefont {K.}~\bibnamefont
  {Kaneko}},\ }\bibfield  {title} {\bibinfo {title} {{Clustering, coding,
  switching, hierarchical ordering, and control in a network of chaotic
  elements}},\ }\href {https://doi.org/10.1016/0167-2789(90)90119-A} {\bibfield
   {journal} {\bibinfo  {journal} {Phys. D}\ }\textbf {\bibinfo {volume}
  {41}},\ \bibinfo {pages} {137} (\bibinfo {year} {1990})}\BibitemShut
  {NoStop}%
\bibitem [{\citenamefont {Golomb}\ \emph {et~al.}(1992)\citenamefont {Golomb},
  \citenamefont {Hansel}, \citenamefont {Shraiman},\ and\ \citenamefont
  {Sompolinsky}}]{Golomb_PRA_1992}%
  \BibitemOpen
  \bibfield  {author} {\bibinfo {author} {\bibfnamefont {D.}~\bibnamefont
  {Golomb}}, \bibinfo {author} {\bibfnamefont {D.}~\bibnamefont {Hansel}},
  \bibinfo {author} {\bibfnamefont {B.}~\bibnamefont {Shraiman}},\ and\
  \bibinfo {author} {\bibfnamefont {H.}~\bibnamefont {Sompolinsky}},\
  }\bibfield  {title} {\bibinfo {title} {{Clustering in globally coupled phase
  oscillators}},\ }\href {https://doi.org/10.1103/PhysRevA.45.3516} {\bibfield
  {journal} {\bibinfo  {journal} {Phys. Rev. A}\ }\textbf {\bibinfo {volume}
  {45}},\ \bibinfo {pages} {3516} (\bibinfo {year} {1992})}\BibitemShut
  {NoStop}%
\bibitem [{\citenamefont {Hakim}\ and\ \citenamefont
  {Rappel}(1992)}]{Hakim_PRA_1992}%
  \BibitemOpen
  \bibfield  {author} {\bibinfo {author} {\bibfnamefont {V.}~\bibnamefont
  {Hakim}}\ and\ \bibinfo {author} {\bibfnamefont {W.~J.}\ \bibnamefont
  {Rappel}},\ }\bibfield  {title} {\bibinfo {title} {{Dynamics of the globally
  coupled complex Ginzburg-Landau equation}},\ }\href
  {https://doi.org/10.1103/PhysRevA.46.R7347} {\bibfield  {journal} {\bibinfo
  {journal} {Phys. Rev. A}\ }\textbf {\bibinfo {volume} {46}},\ \bibinfo
  {pages} {R7347} (\bibinfo {year} {1992})}\BibitemShut {NoStop}%
\bibitem [{\citenamefont {Okuda}(1993)}]{Okuda_PhysicaD_1993}%
  \BibitemOpen
  \bibfield  {author} {\bibinfo {author} {\bibfnamefont {K.}~\bibnamefont
  {Okuda}},\ }\bibfield  {title} {\bibinfo {title} {{Variety and generality of
  clustering in globally coupled oscillators}},\ }\href
  {https://doi.org/10.1016/0167-2789(93)90121-G} {\bibfield  {journal}
  {\bibinfo  {journal} {Phys. D}\ }\textbf {\bibinfo {volume} {63}},\ \bibinfo
  {pages} {424} (\bibinfo {year} {1993})}\BibitemShut {NoStop}%
\bibitem [{\citenamefont {Nakagawa}\ and\ \citenamefont
  {Kuramoto}(1993)}]{Nakagawa_PTPS_1993}%
  \BibitemOpen
  \bibfield  {author} {\bibinfo {author} {\bibfnamefont {N.}~\bibnamefont
  {Nakagawa}}\ and\ \bibinfo {author} {\bibfnamefont {Y.}~\bibnamefont
  {Kuramoto}},\ }\bibfield  {title} {\bibinfo {title} {{Collective Chaos in a
  Population of Globally Coupled Oscillators}},\ }\href
  {https://doi.org/10.1143/ptp/89.2.313} {\bibfield  {journal} {\bibinfo
  {journal} {Prog. Theor. Phys.}\ }\textbf {\bibinfo {volume} {89}},\ \bibinfo
  {pages} {313} (\bibinfo {year} {1993})}\BibitemShut {NoStop}%
\bibitem [{\citenamefont {Kuramoto}\ and\ \citenamefont
  {Battogtokh}(2002)}]{Kuramoto_NPCS_2002}%
  \BibitemOpen
  \bibfield  {author} {\bibinfo {author} {\bibfnamefont {Y.}~\bibnamefont
  {Kuramoto}}\ and\ \bibinfo {author} {\bibfnamefont {D.}~\bibnamefont
  {Battogtokh}},\ }\bibfield  {title} {\bibinfo {title} {{Coexistence of
  Coherence and Incoherence in Nonlocally Coupled Phase Oscillators}},\ }\href
  {http://www.j-npcs.org/abstracts/vol2002/v5no4/v5no4p380.html} {\bibfield
  {journal} {\bibinfo  {journal} {Nonlinear Phenom. Complex Syst.}\ }\textbf
  {\bibinfo {volume} {5}},\ \bibinfo {pages} {380} (\bibinfo {year}
  {2002})}\BibitemShut {NoStop}%
\bibitem [{\citenamefont {Abrams}\ and\ \citenamefont
  {Strogatz}(2004)}]{Abrams_PRL_2004}%
  \BibitemOpen
  \bibfield  {author} {\bibinfo {author} {\bibfnamefont {D.~M.}\ \bibnamefont
  {Abrams}}\ and\ \bibinfo {author} {\bibfnamefont {S.~H.}\ \bibnamefont
  {Strogatz}},\ }\bibfield  {title} {\bibinfo {title} {{Chimera states for
  coupled oscillators}},\ }\href
  {https://doi.org/10.1103/PhysRevLett.93.174102} {\bibfield  {journal}
  {\bibinfo  {journal} {Phys. Rev. Lett.}\ }\textbf {\bibinfo {volume} {93}},\
  \bibinfo {pages} {174102} (\bibinfo {year} {2004})}\BibitemShut {NoStop}%
\bibitem [{\citenamefont {Panaggio}\ and\ \citenamefont
  {Abrams}(2015)}]{Panaggio_Nonlinearity_2015}%
  \BibitemOpen
  \bibfield  {author} {\bibinfo {author} {\bibfnamefont {M.~J.}\ \bibnamefont
  {Panaggio}}\ and\ \bibinfo {author} {\bibfnamefont {D.~M.}\ \bibnamefont
  {Abrams}},\ }\bibfield  {title} {\bibinfo {title} {{Chimera states:
  Coexistence of coherence and incoherence in networks of coupled
  oscillators}},\ }\href {https://doi.org/10.1088/0951-7715/28/3/R67}
  {\bibfield  {journal} {\bibinfo  {journal} {Nonlinearity}\ }\textbf {\bibinfo
  {volume} {28}},\ \bibinfo {pages} {R67} (\bibinfo {year} {2015})}\BibitemShut
  {NoStop}%
\bibitem [{\citenamefont {Sch{\"{o}}ll}(2016)}]{Scholl_EPJST_2016}%
  \BibitemOpen
  \bibfield  {author} {\bibinfo {author} {\bibfnamefont {E.}~\bibnamefont
  {Sch{\"{o}}ll}},\ }\bibfield  {title} {\bibinfo {title} {{Synchronization
  patterns and chimera states in complex networks: Interplay of topology and
  dynamics}},\ }\href {https://doi.org/10.1140/epjst/e2016-02646-3} {\bibfield
  {journal} {\bibinfo  {journal} {Eur. Phys. J. Spec. Top.}\ }\textbf {\bibinfo
  {volume} {225}},\ \bibinfo {pages} {891} (\bibinfo {year}
  {2016})}\BibitemShut {NoStop}%
\bibitem [{\citenamefont {Omel'chenko}(2018)}]{OmelChenko_Nonlinearity_2018}%
  \BibitemOpen
  \bibfield  {author} {\bibinfo {author} {\bibfnamefont {O.~E.}\ \bibnamefont
  {Omel'chenko}},\ }\bibfield  {title} {\bibinfo {title} {{The mathematics
  behind chimera states}},\ }\href {https://doi.org/10.1088/1361-6544/aaaa07}
  {\bibfield  {journal} {\bibinfo  {journal} {Nonlinearity}\ }\textbf {\bibinfo
  {volume} {31}},\ \bibinfo {pages} {R121} (\bibinfo {year}
  {2018})}\BibitemShut {NoStop}%
\bibitem [{\citenamefont {Ashwin}\ and\ \citenamefont
  {Burylko}(2015)}]{Ashwin_Chaos_2015}%
  \BibitemOpen
  \bibfield  {author} {\bibinfo {author} {\bibfnamefont {P.}~\bibnamefont
  {Ashwin}}\ and\ \bibinfo {author} {\bibfnamefont {O.}~\bibnamefont
  {Burylko}},\ }\bibfield  {title} {\bibinfo {title} {{Weak chimeras in minimal
  networks of coupled phase oscillators}},\ }\href
  {https://doi.org/10.1063/1.4905197} {\bibfield  {journal} {\bibinfo
  {journal} {Chaos}\ }\textbf {\bibinfo {volume} {25}},\ \bibinfo {pages}
  {013106} (\bibinfo {year} {2015})}\BibitemShut {NoStop}%
\bibitem [{\citenamefont {Panaggio}\ \emph {et~al.}(2016)\citenamefont
  {Panaggio}, \citenamefont {Abrams}, \citenamefont {Ashwin},\ and\
  \citenamefont {Laing}}]{Panaggio_PRE_2016}%
  \BibitemOpen
  \bibfield  {author} {\bibinfo {author} {\bibfnamefont {M.~J.}\ \bibnamefont
  {Panaggio}}, \bibinfo {author} {\bibfnamefont {D.~M.}\ \bibnamefont
  {Abrams}}, \bibinfo {author} {\bibfnamefont {P.}~\bibnamefont {Ashwin}},\
  and\ \bibinfo {author} {\bibfnamefont {C.~R.}\ \bibnamefont {Laing}},\
  }\bibfield  {title} {\bibinfo {title} {{Chimera states in networks of phase
  oscillators: The case of two small populations}},\ }\href
  {https://doi.org/10.1103/PhysRevE.93.012218} {\bibfield  {journal} {\bibinfo
  {journal} {Phys. Rev. E}\ }\textbf {\bibinfo {volume} {93}},\ \bibinfo
  {pages} {012218} (\bibinfo {year} {2016})}\BibitemShut {NoStop}%
\bibitem [{\citenamefont {Hart}\ \emph {et~al.}(2016)\citenamefont {Hart},
  \citenamefont {Bansal}, \citenamefont {Murphy},\ and\ \citenamefont
  {Roy}}]{Hart_Chaos_2016}%
  \BibitemOpen
  \bibfield  {author} {\bibinfo {author} {\bibfnamefont {J.~D.}\ \bibnamefont
  {Hart}}, \bibinfo {author} {\bibfnamefont {K.}~\bibnamefont {Bansal}},
  \bibinfo {author} {\bibfnamefont {T.~E.}\ \bibnamefont {Murphy}},\ and\
  \bibinfo {author} {\bibfnamefont {R.}~\bibnamefont {Roy}},\ }\bibfield
  {title} {\bibinfo {title} {{Experimental observation of chimera and cluster
  states in a minimal globally coupled network}},\ }\href
  {https://doi.org/10.1063/1.4953662} {\bibfield  {journal} {\bibinfo
  {journal} {Chaos}\ }\textbf {\bibinfo {volume} {26}},\ \bibinfo {pages}
  {094801} (\bibinfo {year} {2016})}\BibitemShut {NoStop}%
\bibitem [{\citenamefont {Dudkowski}\ \emph {et~al.}(2016)\citenamefont
  {Dudkowski}, \citenamefont {Grabski}, \citenamefont {Wojewoda}, \citenamefont
  {Perlikowski}, \citenamefont {Maistrenko},\ and\ \citenamefont
  {Kapitaniak}}]{Dudkowski_SciRep_2016}%
  \BibitemOpen
  \bibfield  {author} {\bibinfo {author} {\bibfnamefont {D.}~\bibnamefont
  {Dudkowski}}, \bibinfo {author} {\bibfnamefont {J.}~\bibnamefont {Grabski}},
  \bibinfo {author} {\bibfnamefont {J.}~\bibnamefont {Wojewoda}}, \bibinfo
  {author} {\bibfnamefont {P.}~\bibnamefont {Perlikowski}}, \bibinfo {author}
  {\bibfnamefont {Y.}~\bibnamefont {Maistrenko}},\ and\ \bibinfo {author}
  {\bibfnamefont {T.}~\bibnamefont {Kapitaniak}},\ }\bibfield  {title}
  {\bibinfo {title} {{Experimental multistable states for small network of
  coupled pendula}},\ }\href {https://doi.org/10.1038/srep29833} {\bibfield
  {journal} {\bibinfo  {journal} {Sci. Rep.}\ }\textbf {\bibinfo {volume}
  {6}},\ \bibinfo {pages} {29833} (\bibinfo {year} {2016})}\BibitemShut
  {NoStop}%
\bibitem [{\citenamefont {R{\"{o}}hm}\ \emph {et~al.}(2016)\citenamefont
  {R{\"{o}}hm}, \citenamefont {B{\"{o}}hm},\ and\ \citenamefont
  {L{\"{u}}dge}}]{Rohm_PRE_2016}%
  \BibitemOpen
  \bibfield  {author} {\bibinfo {author} {\bibfnamefont {A.}~\bibnamefont
  {R{\"{o}}hm}}, \bibinfo {author} {\bibfnamefont {F.}~\bibnamefont
  {B{\"{o}}hm}},\ and\ \bibinfo {author} {\bibfnamefont {K.}~\bibnamefont
  {L{\"{u}}dge}},\ }\bibfield  {title} {\bibinfo {title} {{Small chimera states
  without multistability in a globally delay-coupled network of four lasers}},\
  }\href {https://doi.org/10.1103/PhysRevE.94.042204} {\bibfield  {journal}
  {\bibinfo  {journal} {Phys. Rev. E}\ }\textbf {\bibinfo {volume} {94}},\
  \bibinfo {pages} {1} (\bibinfo {year} {2016})}\BibitemShut {NoStop}%
\bibitem [{\citenamefont {Kemeth}\ \emph {et~al.}(2018)\citenamefont {Kemeth},
  \citenamefont {Haugland},\ and\ \citenamefont {Krischer}}]{Kemeth_PRL_2018}%
  \BibitemOpen
  \bibfield  {author} {\bibinfo {author} {\bibfnamefont {F.~P.}\ \bibnamefont
  {Kemeth}}, \bibinfo {author} {\bibfnamefont {S.~W.}\ \bibnamefont
  {Haugland}},\ and\ \bibinfo {author} {\bibfnamefont {K.}~\bibnamefont
  {Krischer}},\ }\bibfield  {title} {\bibinfo {title} {{Symmetries of Chimera
  States}},\ }\href {https://doi.org/10.1103/PhysRevLett.120.214101} {\bibfield
   {journal} {\bibinfo  {journal} {Phys. Rev. Lett.}\ }\textbf {\bibinfo
  {volume} {120}},\ \bibinfo {pages} {214101} (\bibinfo {year}
  {2018})}\BibitemShut {NoStop}%
\bibitem [{\citenamefont {Wojewoda}\ \emph {et~al.}(2016)\citenamefont
  {Wojewoda}, \citenamefont {Czolczynski}, \citenamefont {Maistrenko},\ and\
  \citenamefont {Kapitaniak}}]{Wojewoda_SciRep_2016}%
  \BibitemOpen
  \bibfield  {author} {\bibinfo {author} {\bibfnamefont {J.}~\bibnamefont
  {Wojewoda}}, \bibinfo {author} {\bibfnamefont {K.}~\bibnamefont
  {Czolczynski}}, \bibinfo {author} {\bibfnamefont {Y.}~\bibnamefont
  {Maistrenko}},\ and\ \bibinfo {author} {\bibfnamefont {T.}~\bibnamefont
  {Kapitaniak}},\ }\bibfield  {title} {\bibinfo {title} {{The smallest chimera
  state for coupled pendula}},\ }\href {https://doi.org/10.1038/srep34329}
  {\bibfield  {journal} {\bibinfo  {journal} {Sci. Rep.}\ }\textbf {\bibinfo
  {volume} {6}},\ \bibinfo {pages} {34329} (\bibinfo {year}
  {2016})}\BibitemShut {NoStop}%
\bibitem [{\citenamefont {Maistrenko}\ \emph {et~al.}(2017)\citenamefont
  {Maistrenko}, \citenamefont {Brezetsky}, \citenamefont {Jaros}, \citenamefont
  {Levchenko},\ and\ \citenamefont {Kapitaniak}}]{Maistrenko_PRE_2017}%
  \BibitemOpen
  \bibfield  {author} {\bibinfo {author} {\bibfnamefont {Y.}~\bibnamefont
  {Maistrenko}}, \bibinfo {author} {\bibfnamefont {S.}~\bibnamefont
  {Brezetsky}}, \bibinfo {author} {\bibfnamefont {P.}~\bibnamefont {Jaros}},
  \bibinfo {author} {\bibfnamefont {R.}~\bibnamefont {Levchenko}},\ and\
  \bibinfo {author} {\bibfnamefont {T.}~\bibnamefont {Kapitaniak}},\ }\bibfield
   {title} {\bibinfo {title} {{Smallest chimera states}},\ }\href
  {https://doi.org/10.1103/PhysRevE.95.010203} {\bibfield  {journal} {\bibinfo
  {journal} {Phys. Rev. E}\ }\textbf {\bibinfo {volume} {95}},\ \bibinfo
  {pages} {010203} (\bibinfo {year} {2017})}\BibitemShut {NoStop}%
\bibitem [{\citenamefont {Schmidt}\ \emph {et~al.}(2014)\citenamefont
  {Schmidt}, \citenamefont {Sch{\"{o}}nleber}, \citenamefont {Krischer},\ and\
  \citenamefont {Garc{\'{i}}a-Morales}}]{Schmidt_Chaos_2014}%
  \BibitemOpen
  \bibfield  {author} {\bibinfo {author} {\bibfnamefont {L.}~\bibnamefont
  {Schmidt}}, \bibinfo {author} {\bibfnamefont {K.}~\bibnamefont
  {Sch{\"{o}}nleber}}, \bibinfo {author} {\bibfnamefont {K.}~\bibnamefont
  {Krischer}},\ and\ \bibinfo {author} {\bibfnamefont {V.}~\bibnamefont
  {Garc{\'{i}}a-Morales}},\ }\bibfield  {title} {\bibinfo {title} {{Coexistence
  of synchrony and incoherence in oscillatory media under nonlinear global
  coupling}},\ }\href {https://doi.org/10.1063/1.4858996} {\bibfield  {journal}
  {\bibinfo  {journal} {Chaos}\ }\textbf {\bibinfo {volume} {24}},\ \bibinfo
  {pages} {013102} (\bibinfo {year} {2014})}\BibitemShut {NoStop}%
\bibitem [{\citenamefont {Sethia}\ and\ \citenamefont
  {Sen}(2014)}]{Sethia_PRL_2014}%
  \BibitemOpen
  \bibfield  {author} {\bibinfo {author} {\bibfnamefont {G.~C.}\ \bibnamefont
  {Sethia}}\ and\ \bibinfo {author} {\bibfnamefont {A.}~\bibnamefont {Sen}},\
  }\bibfield  {title} {\bibinfo {title} {{Chimera states: The existence
  criteria revisited}},\ }\href
  {https://doi.org/10.1103/PhysRevLett.112.144101} {\bibfield  {journal}
  {\bibinfo  {journal} {Phys. Rev. Lett.}\ }\textbf {\bibinfo {volume} {112}},\
  \bibinfo {pages} {144101} (\bibinfo {year} {2014})}\BibitemShut {NoStop}%
\bibitem [{\citenamefont {B{\"{o}}hm}\ \emph {et~al.}(2015)\citenamefont
  {B{\"{o}}hm}, \citenamefont {Zakharova}, \citenamefont {Sch{\"{o}}ll},\ and\
  \citenamefont {L{\"{u}}dge}}]{Bohm_PRE_2015}%
  \BibitemOpen
  \bibfield  {author} {\bibinfo {author} {\bibfnamefont {F.}~\bibnamefont
  {B{\"{o}}hm}}, \bibinfo {author} {\bibfnamefont {A.}~\bibnamefont
  {Zakharova}}, \bibinfo {author} {\bibfnamefont {E.}~\bibnamefont
  {Sch{\"{o}}ll}},\ and\ \bibinfo {author} {\bibfnamefont {K.}~\bibnamefont
  {L{\"{u}}dge}},\ }\bibfield  {title} {\bibinfo {title} {{Amplitude-phase
  coupling drives chimera states in globally coupled laser networks}},\ }\href
  {https://doi.org/10.1103/PhysRevE.91.040901} {\bibfield  {journal} {\bibinfo
  {journal} {Phys. Rev. E}\ }\textbf {\bibinfo {volume} {91}},\ \bibinfo
  {pages} {040901} (\bibinfo {year} {2015})}\BibitemShut {NoStop}%
\bibitem [{\citenamefont {Moehlis}\ and\ \citenamefont
  {Knobloch}(2007)}]{Moehlis_ScholPed_2007}%
  \BibitemOpen
  \bibfield  {author} {\bibinfo {author} {\bibfnamefont {J.}~\bibnamefont
  {Moehlis}}\ and\ \bibinfo {author} {\bibfnamefont {E.}~\bibnamefont
  {Knobloch}},\ }\bibfield  {title} {\bibinfo {title} {{Equivariant bifurcation
  theory}},\ }\href {https://doi.org/10.4249/scholarpedia.2511} {\bibfield
  {journal} {\bibinfo  {journal} {Scholarpedia}\ }\textbf {\bibinfo {volume}
  {2}},\ \bibinfo {pages} {2511} (\bibinfo {year} {2007})}\BibitemShut
  {NoStop}%
\bibitem [{\citenamefont {Kuramoto}(1984)}]{Kuramoto_Book_1984}%
  \BibitemOpen
  \bibfield  {author} {\bibinfo {author} {\bibfnamefont {Y.}~\bibnamefont
  {Kuramoto}},\ }\href {https://doi.org/10.1007/978-3-642-69689-3} {\emph
  {\bibinfo {title} {{Chemical Oscillations, Waves, and Turbulence}}}},\
  \bibinfo {series} {Springer Series in Synergetics}, Vol.~\bibinfo {volume}
  {19}\ (\bibinfo  {publisher} {Springer Berlin Heidelberg},\ \bibinfo
  {address} {Berlin, Heidelberg},\ \bibinfo {year} {1984})\BibitemShut
  {NoStop}%
\bibitem [{\citenamefont {Schmidt}\ and\ \citenamefont
  {Krischer}(2014)}]{Schmidt_PRE_2014}%
  \BibitemOpen
  \bibfield  {author} {\bibinfo {author} {\bibfnamefont {L.}~\bibnamefont
  {Schmidt}}\ and\ \bibinfo {author} {\bibfnamefont {K.}~\bibnamefont
  {Krischer}},\ }\bibfield  {title} {\bibinfo {title} {{Two-cluster solutions
  in an ensemble of generic limit-cycle oscillators with periodic self-forcing
  via the mean-field}},\ }\href {https://doi.org/10.1103/PhysRevE.90.042911}
  {\bibfield  {journal} {\bibinfo  {journal} {Phys. Rev. E}\ }\textbf {\bibinfo
  {volume} {90}},\ \bibinfo {pages} {042911} (\bibinfo {year}
  {2014})}\BibitemShut {NoStop}%
\bibitem [{\citenamefont {Schmidt}\ and\ \citenamefont
  {Krischer}(2015)}]{Schmidt_PRL_2015}%
  \BibitemOpen
  \bibfield  {author} {\bibinfo {author} {\bibfnamefont {L.}~\bibnamefont
  {Schmidt}}\ and\ \bibinfo {author} {\bibfnamefont {K.}~\bibnamefont
  {Krischer}},\ }\bibfield  {title} {\bibinfo {title} {{Clustering as a
  prerequisite for chimera states in globally coupled systems}},\ }\href
  {https://doi.org/10.1103/PhysRevLett.114.034101} {\bibfield  {journal}
  {\bibinfo  {journal} {Phys. Rev. Lett.}\ }\textbf {\bibinfo {volume} {114}},\
  \bibinfo {pages} {034101} (\bibinfo {year} {2015})}\BibitemShut {NoStop}%
\bibitem [{\citenamefont {Haugland}\ and\ \citenamefont
  {Krischer}(2021)}]{Haugland_NatComm_2021}%
  \BibitemOpen
  \bibfield  {author} {\bibinfo {author} {\bibfnamefont {S.~W.}\ \bibnamefont
  {Haugland}}\ and\ \bibinfo {author} {\bibfnamefont {K.}~\bibnamefont
  {Krischer}},\ }\bibfield  {title} {\bibinfo {title} {{A hierarchy of
  coexistence patterns mediating between low- and high-dimensional dynamics in
  highly symmetric systems}},\ }\href {http://arxiv.org/abs/2101.10242}
  {\bibfield  {journal} {\bibinfo  {journal} {arXiv:2101.10242}\ } (\bibinfo
  {year} {2021})}\BibitemShut {NoStop}%
\bibitem [{\citenamefont {Hoyle}(2006)}]{Hoyle_Book_2006}%
  \BibitemOpen
  \bibfield  {author} {\bibinfo {author} {\bibfnamefont {R.~B.}\ \bibnamefont
  {Hoyle}},\ }\href {https://doi.org/10.1017/CBO9780511616051} {\emph {\bibinfo
  {title} {Pattern Formation An Introduction to Methods}}},\ \bibinfo {edition}
  {1st}\ ed.\ (\bibinfo  {publisher} {Cambridge University Press},\ \bibinfo
  {address} {Cambridge},\ \bibinfo {year} {2006})\BibitemShut {NoStop}%
\bibitem [{\citenamefont {Doedel}\ and\ \citenamefont
  {Oldeman}(2019)}]{Doedel_Report_2012}%
  \BibitemOpen
  \bibfield  {author} {\bibinfo {author} {\bibfnamefont {E.}~\bibnamefont
  {Doedel}}\ and\ \bibinfo {author} {\bibfnamefont {B.}~\bibnamefont
  {Oldeman}},\ }\href {http://github.com/auto-07p/auto-07p} {\emph {\bibinfo
  {title} {{Auto 07p: Continuation and bifurcation software for ordinary
  differential equations}}}},\ \bibinfo {type} {Tech. Rep.}\ (\bibinfo
  {institution} {Concordia University and McGill HPC Centre},\ \bibinfo
  {address} {Montreal, Canada},\ \bibinfo {year} {2019})\BibitemShut {NoStop}%
\bibitem [{\citenamefont {Chossat}\ and\ \citenamefont
  {Golubitsky}(1988)}]{Chossat_PhysicaD_1988}%
  \BibitemOpen
  \bibfield  {author} {\bibinfo {author} {\bibfnamefont {P.}~\bibnamefont
  {Chossat}}\ and\ \bibinfo {author} {\bibfnamefont {M.}~\bibnamefont
  {Golubitsky}},\ }\bibfield  {title} {\bibinfo {title} {{Symmetry-increasing
  bifurcation of chaotic attractors}},\ }\href
  {https://doi.org/10.1016/0167-2789(88)90066-8} {\bibfield  {journal}
  {\bibinfo  {journal} {Phys. D}\ }\textbf {\bibinfo {volume} {32}},\ \bibinfo
  {pages} {423} (\bibinfo {year} {1988})}\BibitemShut {NoStop}%
\bibitem [{\citenamefont {Abrams}\ \emph {et~al.}(2008)\citenamefont {Abrams},
  \citenamefont {Mirollo}, \citenamefont {Strogatz},\ and\ \citenamefont
  {Wiley}}]{Abrams_PRL_2008}%
  \BibitemOpen
  \bibfield  {author} {\bibinfo {author} {\bibfnamefont {D.~M.}\ \bibnamefont
  {Abrams}}, \bibinfo {author} {\bibfnamefont {R.}~\bibnamefont {Mirollo}},
  \bibinfo {author} {\bibfnamefont {S.~H.}\ \bibnamefont {Strogatz}},\ and\
  \bibinfo {author} {\bibfnamefont {D.~A.}\ \bibnamefont {Wiley}},\ }\bibfield
  {title} {\bibinfo {title} {{Solvable model for chimera states of coupled
  oscillators}},\ }\href {https://doi.org/10.1103/PhysRevLett.101.084103}
  {\bibfield  {journal} {\bibinfo  {journal} {Phys. Rev. Lett.}\ }\textbf
  {\bibinfo {volume} {101}},\ \bibinfo {pages} {84103} (\bibinfo {year}
  {2008})}\BibitemShut {NoStop}%
\end{thebibliography}%

\end{document}

\preprint{AIP/123-QED}

\title[]{Supplemental material: Connecting minimal chimeras and fully asymmetric chaotic attractors through equivariant pitchfork bifurcations}

\author{Sindre W. Haugland}
\author{Katharina Krischer}
 \email{krischer@tum.de.}
\affiliation{Physics Department, Nonequilibrium Chemical Physics, Technical University of Munich,
  James-Franck-Str. 1, D-85748 Garching, Germany}

% \homepage{http://www.Second.institution.edu/~Charlie.Author.}
%\affiliation{%
%Second institution and/or address%\\This line break forced% with \\
%}%

\date{\today}% It is always \today, today,

\maketitle             

\begin{figure*}[htp]
    \centering
    \includegraphics[width=0.99\textwidth]{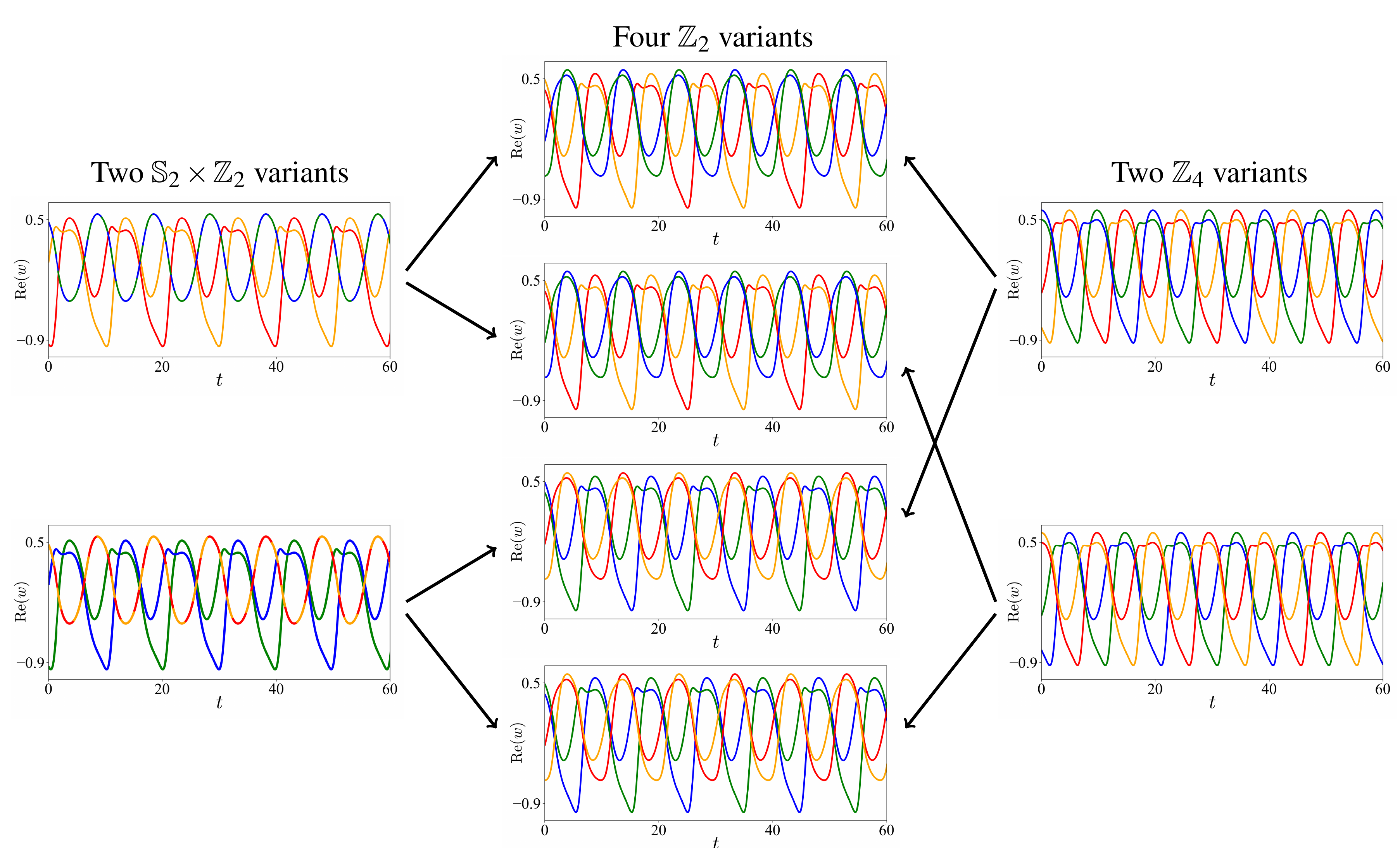}
    \caption{
    Overview of the two variants of the $\mathbb{S}_2\times\mathbb{Z}_2$ (left column) and $\mathbb{Z}_4$ (right column) solutions emerging from one variant of the $2\!-\!2$ solution, as well as the four variants of the intermediate $\mathbb{Z}_2$ solution connecting these.
    Solutions are represented by the time series of the real part of each oscillator in the rotating frame.
    Black arrows indicate which $\mathbb{Z}_2$ variants emerge from each of the variants of the more symmetric solutions.
    Notably, the two $\mathbb{Z}_2$ variants emerging from each $\mathbb{S}_2\times\mathbb{Z}$ variant do not merge into the same $\mathbb{Z}_4$ variant.
    The states shown here are all descended from the $2\!-\!2$ variant wherein the red and the yellow, and the green and the blue oscillator are clustered, respectively.
    The situation is analogous for %the states emerging from 
    each of the two other variants of the $2\!-\!2$~solution.
    }
    \label{fig:N4_2-2_period-doubled_states_ALL_variants}
\end{figure*}

\begin{figure*}[htp]
    \centering
    \includegraphics[width=0.99\textwidth]{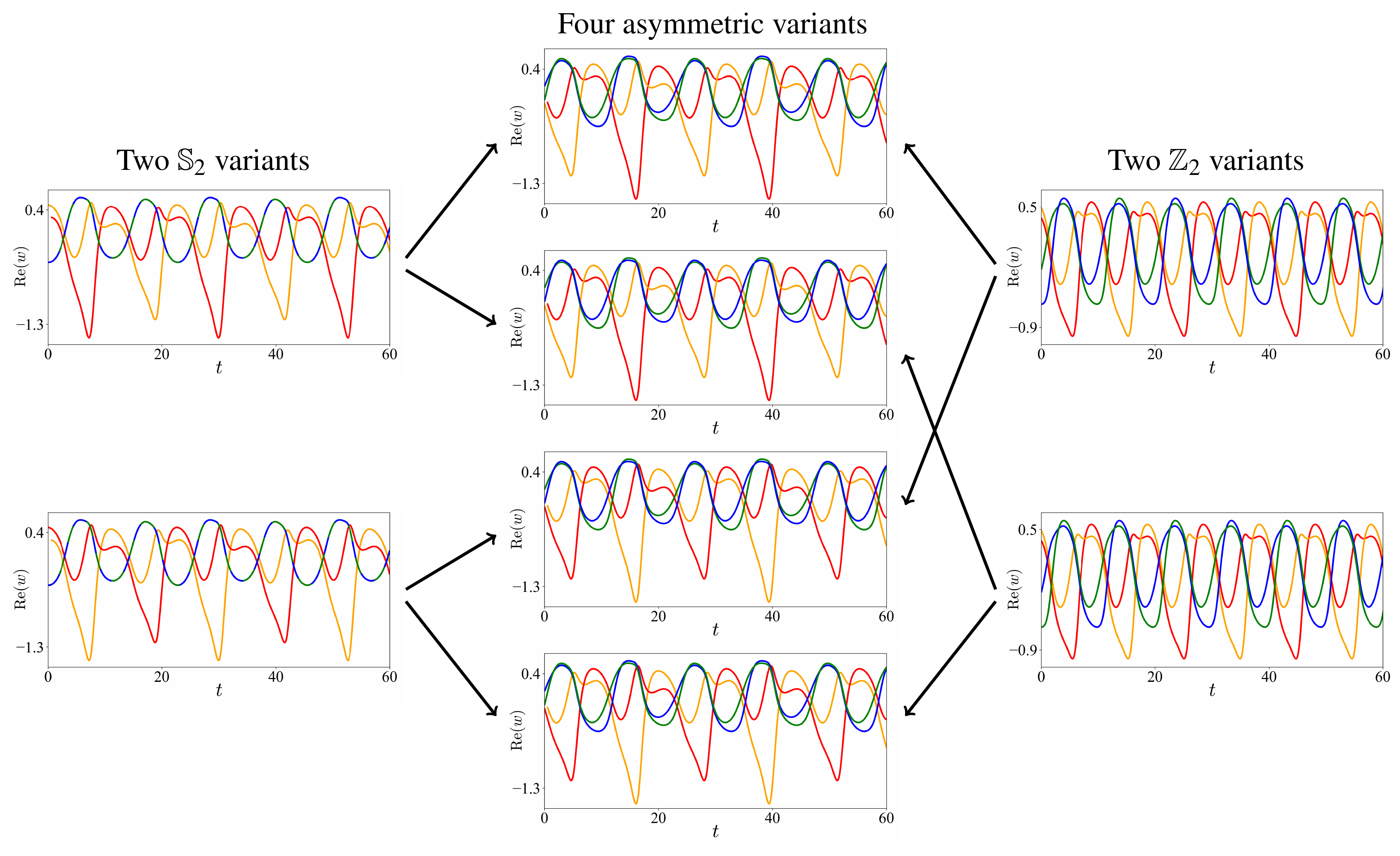}
    \caption{
    Overview of the two variants of the $\mathbb{S}_2$ (left column) and $\mathbb{Z}_2$ (right column) solutions emerging from one variant of the $\mathbb{S}_2\times\mathbb{Z}_2$ solution, as well as the four variants of the fully asymmetric state mediating between these.
    Solutions are represented by the time series of the real part of each oscillator in the rotating frame.
    Black arrows indicate which asymmetric variants emerge from which variants of the more symmetric solutions.
    Notably, the two totally asymmetric variants emerging from each $\mathbb{S}_2$ variant do not merge into the same $\mathbb{Z}_2$ variant.
    The variants shown here are all descended from that variant of the $\mathbb{S}_2\times\mathbb{Z}_2$ solution wherein the green and blue oscillator are clustered,
    that is, the upper left variant in Fig.~\ref{fig:N4_2-2_period-doubled_states_ALL_variants}.
    The situation is analogous for the states emerging from each of the five other variants of the $\mathbb{S}_2\times\mathbb{Z}_2$~solution.
    }
    \label{fig:N4_2-2_period-doubled_states_less_symmetry_ALL_variants}
\end{figure*}

\bibliography{references_manually_edited.bib}